\documentclass[iop]{emulateapj}
\usepackage{color}
\usepackage{graphicx}
\usepackage{amsmath}
\usepackage{threeparttable}

\newcommand{\Mpc}{\mathrm{~km~s^{-1}~Mpc^{-1}}}

\shorttitle{Direct tests of general relativity under screening effect}
\shortauthors{Lian, et al.}

\begin{document}
\title{Direct tests of General Relativity under screening effect with galaxy-scale strong lensing systems}
\author{Yujie Lian\altaffilmark{1,2}, Shuo Cao\altaffilmark{1,2$\ast$}, Tonghua Liu\altaffilmark{3}, Marek Biesiada\altaffilmark{4}, and Zong-Hong Zhu\altaffilmark{1,2}}

\altaffiltext{1}{Institute for Frontiers in Astronomy and Astrophysics, Beijing Normal University, Beijing 102206, China; \emph{caoshuo@bnu.edu.cn}}
\altaffiltext{2}{Department of Astronomy, Beijing Normal University, Beijing 100875, China;}
\altaffiltext{3}{School of Physics and Optoelectronic, Yangtze University, Jingzhou 434023, China;}
\altaffiltext{4}{National Centre for Nuclear Research, Pasteura 7, 02-093 Warsaw, Poland}

\begin{abstract}

Observations of galaxy-scale strong gravitational lensing (SGL) systems have enabled unique tests of nonlinear departures from general relativity (GR) on the galactic and supergalactic scales. One of the most important cases of such tests is constraints on the gravitational slip between two scalar gravitational potentials. In this paper, we use a newly compiled sample of strong gravitational lenses to test the validity of GR, focusing on the screening effects on the apparent positions of lensed sources relative to the GR predictions. This is the first simultaneous measurement of the Post-Newtonian (PN) parameter ($\gamma_{PN}$) and the screening radius ($\Lambda$) without any assumptions about the contents of the Universe. Our results suggest that the measured PPN is marginally consistent with GR ($\gamma_{PN}=1$) with increasing screening radius ($\Lambda = 10-300 $kpc), although the choice of lens models may have a significant influence on the final measurements. Based on a well-defined sample of 5000 simulated strong lenses from the forthcoming LSST, our methodology will provide a strong extragalactic test of GR with the accuracy of 0.5\%, assessed up to scales of $\Lambda \sim 300$ kpc. For the current and future observations of available SGL systems, there is no noticeable evidence indicating some specific cutoff scales on kpc-Mpc scales, beyond which new gravitational degrees of freedom are expressed.

\end{abstract}


\section{Introduction} \label{introduction}

As a successful theory of a dynamical space time where general coordinate invariance acts an essential role, Einstein's Theory of General Relativity (GR) has passed all observational test so far \citep{Ashby02,Bertotti03}, from the millimetre scale in the laboratory to the Solar System tests and consistency with gravitational wave emission by binary pulsars \citep{Dyson20,Pound60,Shapiro64,Taylor79,Shapiro04,Williams04}. A recent review about the status of experimental tests of GR and of theoretical frameworks for analyzing them can be found in \citet{Will2014}. Extrapolating to the cosmological scales, GR seems to precisely characterize the history  and evolution of the Universe, as well as the large scale structure of space-time and matter. On the other hand, a mysterious component with negative pressure, dubbed dark energy (DE) \citep{Copeland2006} responsible for the accelerating expansion of the Universe should be invoked in the framework of GR, and the simplest candidate of DE is the cosmological constant (introduced for different reasons at the early years of the GR). However, the inconsistency between the observed value of this constant and the theoretical value of zero-point energy predicted by quantum field theory is considerable \citep{Weinberg89}, which is a fairly challenging problem in theoretical physics and opens up a discussion of whether the GR could fail at larger, cosmological scales \citep{Koyama2016}.
Therefore, besides presenting possible solutions to the cosmological constant problem \citep{Sahni2000,Padmanabhan03,Peebles03,Mukohyama2004,Nojiri2006,Padilla2015}, it would be equally important to formulate and quantitatively interpret the tests of GR on the extra-galactic scale with high precision, and the parameterized post-Newtonian (PPN) framework \citep{Thorne71} offers an interesting approach to test the departure from GR. The post-Newtonian parameter denoted by $\gamma_{PN}$, which measures
the amount of spatial curvature per unit mass, can be used to probe the deviations from GR (i.e. from the value $\gamma_{PN}=1$) on significantly larger length scales due to the scale independence.

Recently, applying time-delay measurements of strongly lensed quasars, \citet{Jyoti2019} proposed a new phenomenological model of gravitational screening as a step discontinuity in the measurement of $\gamma_{PN}$ at a cutoff scale $\Lambda$ to test the nonlinear departure from GR. In this model, \citet{Jyoti2019} made use of two key characteristics of most modified gravity (MG) theories: gravitational slip, meaning the difference between the gravitational potential created by temporal and spatial metric components \citep{Daniel08}, and screening, which can restore GR on small scales but may lead to distinct signatures in the large scale structure of the universe  \citep{Vainshtein72,Jain2010,Joyce2015,Ferreira19}. It is worthy to note that the constraint given by \citet{Jyoti2019} is $|\gamma_{\rm PN}-1| \leq 0.2 \times (\Lambda/100\,\rm kpc)$, with screening length $\Lambda = 10-200$ kpc, which is limited by $10^{-1} \sim 10^{-2}$ precision of the strong lensing time delay measurements using quasars. More recently, \citep{Abadi2021} proposed to use strongly lensed fast radio burst time-delay measurements to put constraints on $\gamma_{PN}$, and yield constraints as tight as $\left | \gamma _{PN}-1 \right |\leqslant 0.04\times (\Lambda /100 \; kpc)\times [N/10]^{-1/2}$ ($N$ denoting the sample size), which indicates that ten events alone could place constraints at a level of 10\% in the range of $\Lambda = 10-300 \; kpc$. Therefore, extra cosmological probes are also required to make complementary and more precise studies of the screened MG model featuring a gravitational slip.

In the last decade, strong gravitational lensing (SGL) of galaxies have become powerful and  promising probes to study astrophysical issues \citep{Treu2010}, such as the structure and  evolution of galaxies, the parameters that characterize the geometry, content, and expansion of the Universe. In recent years, great efforts have been made in estimating cosmological parameters through SGL \citep{Zhu00,Chae2003,Cha04,Mit05,Grillo08,Oguri08,Zhu08a,Zhu08b, Cao12a,Cao12b,Cao12c,Cao12d,Biesiada06,Biesiada10,Collett14,Cardone2016,Amante2020}, in measuring the Hubble constant $H_{0}$ \citep{Fassnacht02,Bonvin16,Suyu2017,Birrer2019,Wong2020},
the cosmic curvature \citep{Rasanen2015,Xia2017,Qi2019,Liu2020,Zhou2020}, and the distribution  of matter in massive galaxies acting as lenses \citep{Zhu97,MS98,Jin00,Kee01,Kochanek2001,Ofek03,Treu2006,Chen2019,Geng2021}.
It should be stressed that all of the studies mentioned above have been carried out under the assumption of GR. On account of some controversies in the framework of GR, for instance the aforementioned cosmological constant problem, and the Hubble tension between different cosmological probes, we are prompted to test GR in this paper. Moreover, SGL by galaxies provide a unique opportunity to test GR in kpc scales
\citep{Bolton2006,Bolton08,Smith2009,Schwab2010,Cao2017,Collett2018,Yang2020,Liu2022}, with
reasonable prior assumptions and independent measurements of background cosmology and appropriate descriptions of the internal structure of lensing galaxies. Inspired by \citet{Schwab2010} and \citet{Cao2017}, who used galaxy-scale SGL systems with measured stellar velocity dispersions to test GR, we utilize the recently combined sample of 158 SGL systems, which is summarized in \citet{Cao2015,Shu2017}. Based on this sample we investigate gravitational slip with screening effect in a phenomenological model. Considering the limitation of the sample size of available SGL systems, we also take advantage of the simulated future LSST SGL data to assess the expected improvement in precision and study the degeneracy between $\gamma_{PN}$ and $\Lambda$.

In this paper, we focus on demonstrating the possibilities of testing gravitational slip  at super-galactic screening scales $\Lambda$ by the observed velocity dispersion of SGL systems.
As already mentioned, our approach to test gravitational slip at super-galactic screening scales $\Lambda$, invokes a combination of lensing and stellar kinematics.  Two different lens models, with a power law density profiles for the total mass density and luminous density (i.e. stellar mass density) assumed, are used to illustrate the influence of the lens mass distribution on the PPN parameter $\gamma_{PN}$. It would be interesting to figure out if the constraints from the current and future SGL systems reveal any specific cutoff scales, beyond which MG is relevant, or GR is still valid with high precision up to scales of $\Lambda \sim 300$ kpc. Following the simplifying assumptions presented by \citep{Jyoti2019}, we also assume that the screening radius $\Lambda$ is bigger than the Einstein radius of the lens galaxy, and thus the stellar dynamics within the lens are not changed, but the departure from GR at larger radius would impact the photon path. This assumption will allow us to investigate the regime of super-galactic screeing, complementary to the discussions in \citep{Bolton2006,Smith2009,Collett2018}, where the screening effects occur within the galaxy and the post-Newtonian parameter $\gamma_{PN}$ is constrained through two different mass measurements: the dynamical mass measured from the stellar kinematics of the deflector galaxy, and the lensing mass inferred from the lensing image.

This paper is organized as follows. In Sec. 2, we give a brief introduction of the model we used to evaluate the velocity dispersion of lensing galaxies. In Sec. 3, all the observational and simulated data sets are  introduced. We perform a Markov chain Monte Carlo (MCMC) analysis using different data sets, and discuss our results in Sec.4. Finally, conclusions are summarized in Sec. 5.

\section{The model}

The post-Newtonian variables are applied to quantify the behavior of gravity and deviations from GR, and we adopt the notation and conventions of \citet{Ma1995} to express the perturbed Robertson-Walker metric
\begin{equation}
ds^2 = a^2(\eta)\left[-(1 + 2 \Phi) d\eta^2 + (1 - 2 \Psi)d \vec x^2\right],
\end{equation}
to characterize the cosmological space-time, where $a(\eta)$ is the cosmological scale factor ($\eta$ being the conformal time), $\Phi$ and
$\Psi$ are Newtonian and longitudinal gravitational potentials. 
In the weak-field limit GR predicts $\Phi = \Psi$ and a gravitational slip manifested as $\Phi \neq \Psi$ occurs in MG theories, for instance scalar-tensor theories \citep{Schimd2005}, $f(R)$ theories \citep{Chiba2003,Sotiriou2010}, Dvali-Gabadadze-Porrati (DGP) model \citep{Dvali2000,Sollerman2009}, and massive gravity \citep{Dubovsky2004}. We remark here that for a modified gravity theory where the Newtonian and longitudinal gravitational potentials are different, i.e. the gravitational slip is not zero, the Poisson equation $\nabla^2\Phi=4\pi G a^2\rho$ usually is modified into a form $\nabla^2\Phi=4\pi G \mu a^2\rho$ which is different from that of GR, where the modified gravity parameters are included in the $\mu$ term \citep{Koyama2016}. Rigorous approach would thus require to focus on a specific theory of modified gravity. Similar to \citet{Jyoti2019} we are taking a heuristic approach assuming that the screening mechanism recovers GR exactly in the central parts of lensing galaxies, so that standard spherical Jeans equation is sufficient to recover the dynamical mass inside the Einstein radius.

The deviation from GR is quantified by the ratio $\gamma_{PN} = \Psi / \Phi$, and $\gamma_{PN} = 1$ represents the GR. Gravitational screening mechanisms can restore GR in regions of high density, potential or curvature, such as galaxies, but allow modification of gravity at cosmological scales, which corresponds to suppressing the additional gravitational degrees of freedom within a certain region. Recent reviews about screening
mechanisms and the experimental tests of such theories are available \citep{Joyce2015,Ferreira19}.
In order to model the effect of screening, we follow the notations in \citet{Jyoti2019} and suppose the gravitational slip
to be a step-wise function, discontinuous at a screening radius $\Lambda$, which is assumed to be larger than the (physical) Einstein radius, $\Lambda>R_E=D_L\theta_E$.

Photon geodesics feel the sum of Newtonian and longitudinal gravitational potentials, which is defined as $\Sigma \equiv \Phi + \Psi$.
Then, for a spherically-systematic mass distribution, the departure from GR can be expressed as
\begin{equation}
\Sigma = [2 + (\gamma_{\rm PN}-1)\Theta(r-\Lambda)]\Phi(r),
\end{equation}
where r and $\Lambda$ are physical distances, and $\Theta$ is the Heaviside step function. For $r \leq \Lambda$, Eq.~(2) reduces to $\Sigma = 2\Phi$ as in GR.
According to the definition in \citet{Narayan1996}, the lensing potential is
\begin{equation}
\psi(\mathbf \theta) = \frac{1}{c^{2}} \frac{D_{LS}}{D_L D_S} \int \Sigma(D_{L} \theta,\mathcal{Z}) d\mathcal{Z}\,
\end{equation}
where $D_{L}$, $D_{S}$, $D_{LS}$ are the angular diameter distances from observer to the lens, the source, and between the
lens and source, and $\mathcal{Z}$ represents the distance along the line of sight. Then, the lensing potential can be decomposed as
\begin{equation}
\psi = \psi_{\rm GR} + (\gamma_{\rm PN}-1) \Delta \psi,
\end{equation}
In this model, $\psi_{\rm GR}$ and $\Delta \psi$, denote the lensing potential in GR and the correction due to screening, which can be expressed as
\begin{equation}
\psi_{GR}(\mathbf \theta) = \frac{2}{c^{2}} \frac{D_{LS}}{D_L D_S} \int \Phi(D_{L} \theta,\mathcal{Z}) d\mathcal{Z}\,
\end{equation}
\begin{equation}
\Delta \psi(\mathbf \theta) = \frac{1}{c^{2}} \frac{D_{LS}}{D_L D_S} \int \Theta(r-\Lambda)\Phi(D_{L} \theta,\mathcal{Z}) d\mathcal{Z}\,
\end{equation}
respectively. Assuming the power-law density profiles for the total mass density for elliptical galaxies,
\begin{equation}
\rho (r) = \rho _{0}\left ( \frac{r}{r_{0}} \right )^{-\gamma}
\end{equation}
where the constants: $\rho_0$ and $r_0$, set the total mass of the lens and solving the Poisson
equation $\nabla^2 \Phi = 4 \pi G a^2 \rho$, one can get the Newtonian potential:
\begin{equation}
\Phi = \frac{4 \pi \rho_{0} r_{0}^{\gamma}}{(\gamma-3)(\gamma-2)}\,r^{2-\gamma}.
\end{equation}
This will lead to the lensing potential in GR \citep{Suyu2012,Jyoti2019,Abadi2021}
\begin{equation}
\psi_{\rm GR} (\theta) = \frac{\theta_{E,\rm GR}^{\gamma-1}}{3-\gamma} ~\theta^{3-\gamma},
\end{equation}
where the $r_{0}$ and $\rho_{0}$ parameters have been subsumed into $\theta_{E,\rm GR}$, and $\theta_{E,\rm GR}$ in
the Einstein radius of the lens inferred within GR. The deflection angle becomes
\begin{equation}
\alpha_{\rm GR} (\theta) = \partial_\theta \psi_{\rm GR}(\theta) =
\theta_{E,\rm GR}^{\gamma-1} \theta^{2-\gamma}.
\end{equation}
One can obtain the PPN correction to the lensing potential $\Psi$ by integrating the potential in Eq.~(6) \citep{Jyoti2019}:
\begin{eqnarray} \Delta\psi(\theta)&=&\frac{c'\, \theta_{E,\rm GR}^{\gamma-1}\, }{3-\gamma} \left(\frac{D_L}{\Lambda}\right)^{\gamma-3}
\\&~&~\times~{}_2F_1\left[\frac{1}{2},\frac{\gamma-3}{2},\frac{\gamma-1}{2};\left(\frac{D_L\theta}{\Lambda}\right)^2\right],\nonumber
\end{eqnarray}
where ${}_2F_1$ is the hypergeometric function,
\begin{equation}
c' = \frac{1 }{2\sqrt{\pi}}~ \frac{\Gamma\left(\frac{\gamma}{2}-1\right)}{\Gamma\left(\frac{\gamma-1}{2}\right)},
\label{eq:const}
\end{equation}
and $\Gamma$ is the Euler's Gamma function. Then, the correction to the deflection angle from Eq.~(11) can be expressed as
$\Delta \alpha = \partial_{\theta} \Delta \psi$.
It should be pointed out that the gravitational slip correction to the lensing potential $\Delta \psi$ influences the lens parameters,
such as time delay, image positions, observational Einstein radius $\Theta_{E,\rm obs}$,
and so on, which are deduced from the lensing observables \citep{Jyoti2019}. For each observed lensing event, one should carry out the Markov chain Monte Carlo (MCMC)
method to analyze the entire set of lens parameters under the effect of $\gamma_{PN}$ at a screening radius $\Lambda$, which is a costly process
for a sample of SGL systems. Therefore, we follow the same procedure as adopted in \citep{Jyoti2019}, relating the unobservable GR Einstein angle $\theta_{E, \rm GR}$
and the observed $\theta_{E,\rm obs}$, which would took into account the gravitational slip, through the lens equation \citep{Narayan1996}
\begin{equation}
\beta ( \theta )=\theta-\alpha_{\rm GR} (\theta)-(\gamma_{PN}-1)\Delta \alpha (\theta),
\end{equation}
where $\beta$, $\theta$, as well as $\alpha$ are the source position, the angular position of lens images, and the deflection angle respectively. Setting
$\beta = 0$ in the equation above, we obtain
\begin{equation}
\theta_{E,\rm obs} = \alpha_{\rm GR}(\theta_{E,\rm obs}) + (\gamma_{PN}-1)\Delta \alpha(\theta_{E,\rm obs} ).
\end{equation}
According to $\Delta \alpha = \partial_{\theta} \Delta \psi$, the exact solution for $\theta_{E,\rm GR}$ is found to be
\begin{equation}
\begin{split}
\theta_{E,\rm GR}=& \left\{\theta_{E,\rm obs}^{1-{\gamma}} - \frac{c'}{2(\gamma-1)} ({\gamma_{\rm PN}}-1) \left(\frac{D_{L}} {\Lambda}\right)^{{\gamma}-1} \right.\\
&\left. \times {}_2F_1 \left[\frac{3}{2},\frac{\gamma-1}{2}; \frac{\gamma+1}{2};\left (\frac{D_{L} \theta_{E,\rm obs}}{\Lambda}  \right )^2 \right] \right\}^{\frac{1}{1-\gamma}},
\end{split}
\end{equation}
where $c'$ is given by Eq.~(12), and $\gamma_{\rm PN}=1$ will lead to $\theta_{E,\rm GR}=\theta_{E,\rm obs}$. After the slip-term is introduced into the GR Einstein angle $\theta_{E,\rm GR}$, we want to express the averaged observed velocity dispersion under the gravitational slip. According to the theory
of gravitational lensing \citep{Schneider92,Schwab2010}, the angular size of the Einstein radius corresponding to a point mass $M_{E}$ (or the spherically symmetric mass distribution within the Einstein radius) is expressed as
\begin{equation}
\theta_{E,\rm GR} =   \left(\frac{4G M(\theta_{E,\rm GR})}{c^2} \frac{D_{LS}}{D_S D_L} \right)^{1/2} ~~~,
\end{equation}
where $M(\theta_{E,\rm GR})$ is the mass enclosed within a radius of $\theta_{E,\rm GR}$ and should not be affected by the modification of GR. We can acquire a useful expression if we rearrange terms in Eq.~(16) with $R_{E}=D_{L} \theta_{E, \rm GR}$ \citep{Schneider92,Abadi2021}:
\begin{equation}
\frac{G M(\theta_{E,\rm GR})}{R_E} = \frac{c^2}{4}
\frac{D_S}{D_{LS}} \theta_{E,\rm GR} ~~~.
\end{equation}

One of the simplest models to describe the mass density of an elliptical galaxy is a scale-free model based on power-law density profiles for the total mass density, $\rho$, and
luminosity density, $\nu$, \citep{Koopmans06a}:
\begin{eqnarray}
\rho(r) &=& \rho_0 \left(\frac{r}{r_0}\right)^{-\gamma} \\
\nu(r) &=& \nu_0 \left(\frac{r}{r_0}\right)^{-\delta}
\end{eqnarray}
Here $r$ is the spherical radial coordinate from the lens center: $r^2 = R^2 + \mathcal{Z}^2$. One can use the anisotropy parameter $\beta$ to characterize anisotropic distribution of
three-dimensional velocity dispersion pattern, which can be written as:
\begin{equation}
\beta(r) = 1 - {\sigma^2_t} / {\sigma^2_r},
\end{equation}
where $\sigma^2_t$ and $\sigma^2_r$ represent the tangential and radial components of the velocity dispersion. In our analysis, $\beta$ is assumed to be independent of $r$, and two cases
will be considered: isotropic dispersion $\beta=0$ and anisotropic dispersion $\beta = const. \neq 0$. Applying the spherical Jeans equation \citep{Binney80}, the radial dispersion of luminous matter $\sigma_r^2(r)$ of the early-type galaxies can be written as
\begin{equation}
\sigma^2_r(r) =  \frac{G\int_r^\infty dr' \ \nu(r') M(r') (r')^{2
		\beta - 2} }{r^{2\beta} \nu(r)}~~~,
\end{equation}
where $\beta$ is the anisotropy parameter. Applying the mass density profile in Eq.~(18), one can obtain the mass contained within a spherical radius $r$ and the total mass $M_{E}$:
\begin{equation}
M(r) = \frac{2}{\sqrt{\pi} \lambda(\gamma)}
\left(\frac{r}{R_E}\right)^{3 - \gamma} M_E ~~~,
\end{equation}
where $\lambda(x) =\Gamma \left(\frac{x-1}{2}\right) / \Gamma \left(\frac{x}{2}\right)$ denotes the ratio of Euler's gamma functions. Following the same formulae with the notation used in \citep{Koopmans06a} : $\xi =\delta+\gamma-2$, we acquire the radial velocity dispersion by scaling the dynamical mass to the Einstein radius:
\begin{equation}
\sigma^2_r(r) = \left[\frac{G M_E}{R_E} \right]
\frac{2}{\sqrt{\pi}\left(\xi- 2 \beta \right) \lambda(\gamma)}
\left(\frac{r}{R_E}\right)^{2 - \gamma}.
\end{equation}
Additionally, \citep{Schwab2010,Cao2016,Cao2017} have pointed out that the actually observed velocity dispersion is measured over the effective spectroscopic aperture $\theta_{ap}$ and effectively averaged over the line-of-sight effects. Given the entanglement of the aperture with atmospheric blurring and luminosity-weighted averaging, the averaged observed velocity dispersion takes the form:
\begin{eqnarray} \label{Schwab}
\nonumber \bar {\sigma}_*^2 &=& \left[\frac{c^2}{4}
\frac{D_s}{D_{ls}} \theta_{E,\rm GR} \right] \frac{2}{\sqrt{\pi}}
\frac{(2 \tilde{\sigma}_{\rm atm}^2/\theta_{E,\rm GR}^2)^{1-\gamma/2}}{ (\xi - 2\beta)} \\
&&\times\left[\frac{\lambda(\xi) - \beta \lambda(\xi+2)}
{\lambda(\gamma)\lambda(\delta)}\right] \frac{
	\Gamma(\frac{3-\xi}{2}) }{\Gamma(\frac{3 - \delta}{2}) } ~~~.
\end{eqnarray}
where $\tilde{\sigma}_{\rm atm}\approx\sigma_{\rm atm} \sqrt{1 + \chi^2 / 4 + \chi^4 / 40}$ and $\chi = \theta_{\rm ap} / \sigma_{\rm atm}$ \citep{Schwab2010}, the detailed expression of $\theta_{E,\rm GR}$ is given by Eq.~(15), $\sigma_{\rm atm}$ is the seeing recorded by the spectroscopic guide cameras during observing sessions, and specific values of seeing for different surveys have been summarized in \citep{Cao2016}. According to Eq.~(24), one can probe the PPN parameter $\gamma_{PN}$ and the screening radius $\Lambda$ on a sample of lenses with available information about measured redshifts of the lens and the source, velocity dispersion, as well as the Einstein radius $\theta_{E,obs}$. In this work, a flat $\Lambda CDM$ model is assumed as a fiducial cosmology with $H_{0}=67.36\Mpc$ and $\Omega_{m}=0.315$ \citep{Planck2018}. In light of some evidence \citep{Koopmans06b,Ruff2011,Bolton2012,Sonnenfeld2013,Cao2016} supporting that the total density profile $\gamma$ for massive galaxies shows a trend of the cosmic evolution, we consider the lens models where the total mass density evolves as a function of redshift for two cases: $\gamma=\delta$ and $\gamma \neq \delta$. Meanwhile, we can examine the possible influence of different lens models on the constraints of $\gamma_{PN}$ and $\Lambda$. As for the anisotropy parameter, $\beta = 0$ is assumed in the case $\gamma=\delta$, and $\beta$ is being marginalized using a Gaussian prior with $\beta=0.18 \pm 0.13$ for the case $\gamma \neq \delta$
\citep{Gerhard2001,Schwab2010,Cao2016}.

\begin{figure}
	\includegraphics[width=\columnwidth]{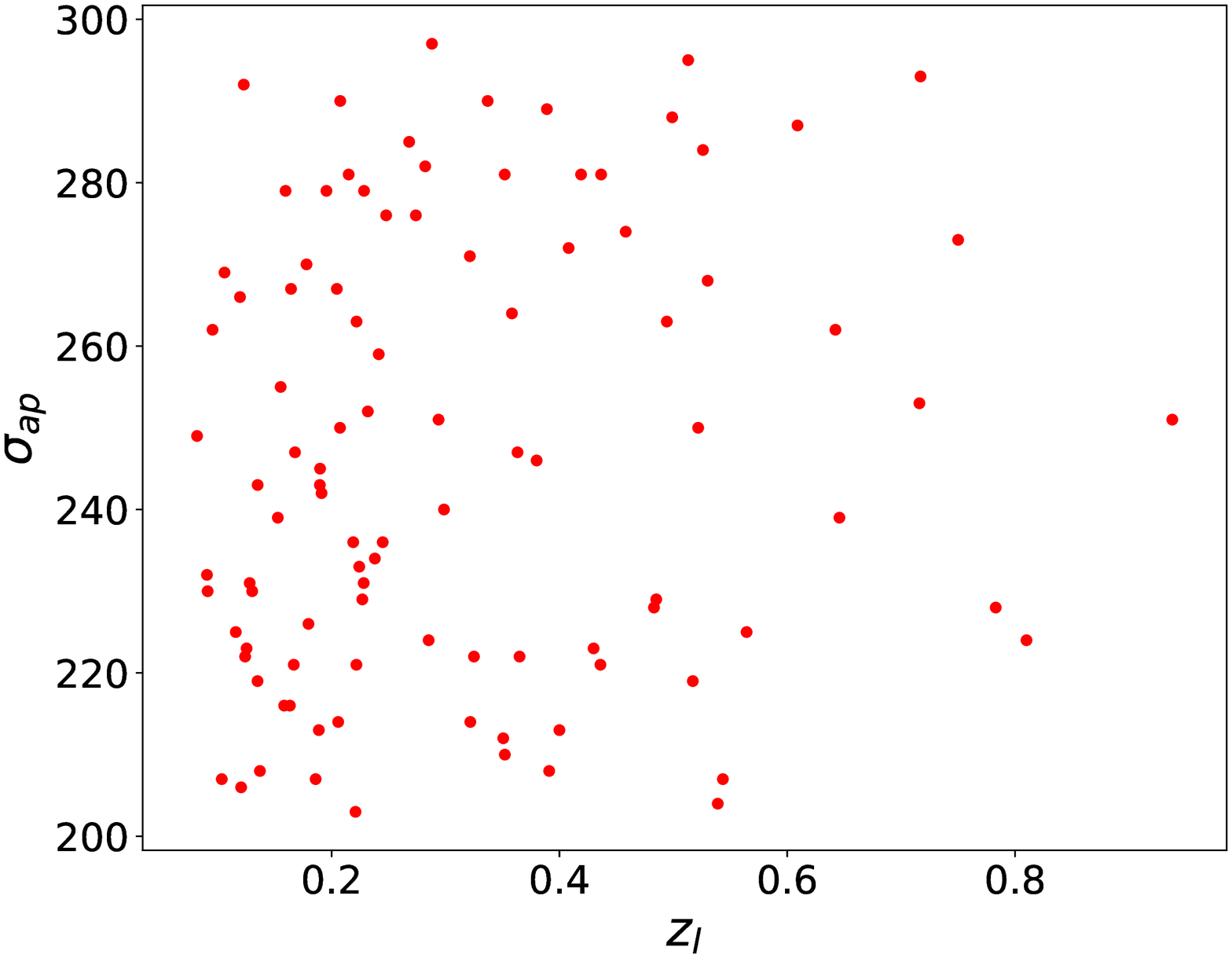}
	\caption{The scatter plot of the current strong lensing sample of 99 intermediate mass early type galaxy. A fair coverage of redshifts can be noted in this combined sample. Velocity dispersion $\sigma_{ap}$ is expressed in ($km/s$).}
\end{figure}

\section{Method and data}

In this section, we present the details of observational and simulated SGL systems
which are used to put constraints on the post-Newtonian slip parameter $\gamma_{PN}$ and the screening radius $\Lambda$.

\subsection{The observational and simulated strong gravitational lensing systems}

Strong gravitational lensing by galaxies has been a powerful tool to probe both astrophysics and cosmology. Moreover, compared with late-type and unknown-type counterparts, early-type galaxies are more probable candidates for intervening lenses for the background sources, because most of the cosmic stellar mass of the Universe is included in such galaxies. \citet{Cao2015} compiled a sample comprising 118 early-type gravitational lenses observed in Sloan Lens ACS Survey (SLACS), the BOSS emission-line lens survey (BELLS), Lenses Structure and Dynamics (LSD), and the Strong Lensing Legacy Survey (SL2S) surveys, which have been used to probe cosmological models, test the fundamental assumptions in cosmology and study the mass density dispersion in early-type galaxies. Applying the 118 SGL sample, \citet{Cao2016} showed that the intermediate-mass lenses ($200 \, km/s< \sigma_{ap} < 300 \, km/s$) are suitable to minimize the possible discrepancy between Einstein mass and dynamical mass for the SIS model. Then \citet{Cao2017} followed this analysis to test the parametrized post-Newtonian gravity with 80 intermediate mass early-type lenses. In this paper, taking into account the sample compiled and summarized in \citet{Cao2015} and \citet{Shu2017}, including 158 early-type SGL systems, we also tend to test GR with a mass-selected sample of SGL systems by restricting the velocity dispersion of lensing galaxies to the intermediate range: $200 \, km/s< \sigma_{ap} < 300 \, km/s$, where 61 lenses are taken from SLACS, 22 lenses from SL2S, 13 lenses from BELLS, and 3 lenses from LSD. Fig.~1 presents the scatter plot for this mass-selected sample.

Strong lensing systems offer a unique opportunity to conduct cosmological research, however, they still suffer from the limited sample size and may not achieve precise enough results on basic cosmological parameters compared with other popular cosmological probes, such as type Ia supernovae (SN Ia), the cosmic microwave background (CMB), and baryon acoustic oscillations (BAO). In the future, the lens sample size is reckoned to increase by orders of magnitude through the next generation of wide and deep sky surveys. Recent studies forecasted the number of galactic-scale lenses that could be discovered in spectroscopic \citep{Serjeant2014} and photometric surveys \citep{Collett2015}, such as the LSST \citep{Abell2009} and the Dark Energy Survey (DES) \citep{Frieman2004}, which are the future wide and deep surveys, broadly expected to revolutionize the strong lensing science by increasing the number of known galactic lenses.
For instance, the forthcoming LSST survey is expected to discover $\sim 10^{5}$ SGL systems \citep{Collett2015} in the near future, and there have been studies to illustrate the performance of the forecasted yield of the LSST in cosmological studies \citep{Cao2017,Cao2018,Qi2019,Ma2019,Cao2020,Liu2020}. It would be promising to extend the current research on the gravitational slip under screening effects to a new regime, that is the ability to detect $\sim 5000$ galaxy-scale lens population in the future LSST survey.

Taking advantage of the publicly available simulation programs \footnote{https://github.com/tcollett/LensPop} elaborately described in \citet{Collett2015}, we have simulated a realistic population of SGL systems with early-type galaxies serving as lenses to anticipate the yields of LSST. Following the appropriate assumptions in simulating procedures \citep{Liu2019}, spherically symmetric power-law distribution is adopted to model the mass distribution of lensing galaxies, meanwhile, the normalization and shape of the velocity dispersion function of early-type galaxies are not changing with redshift. These assumptions are in good agreement with the previous studies on lensing statistics \citep{Chae2003}.

\begin{figure}
	\includegraphics[width=\columnwidth]{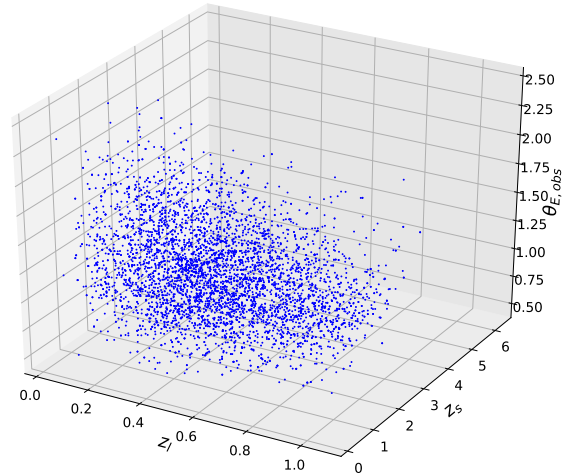}
	\caption{The scatter plot of the simulated LSST forthcoming samples with 5000 SGL systems. Einstein radius is given in $arcsec$.}
\end{figure}

\begin{figure*}
	\begin{center}
		\includegraphics[width=0.6\linewidth]{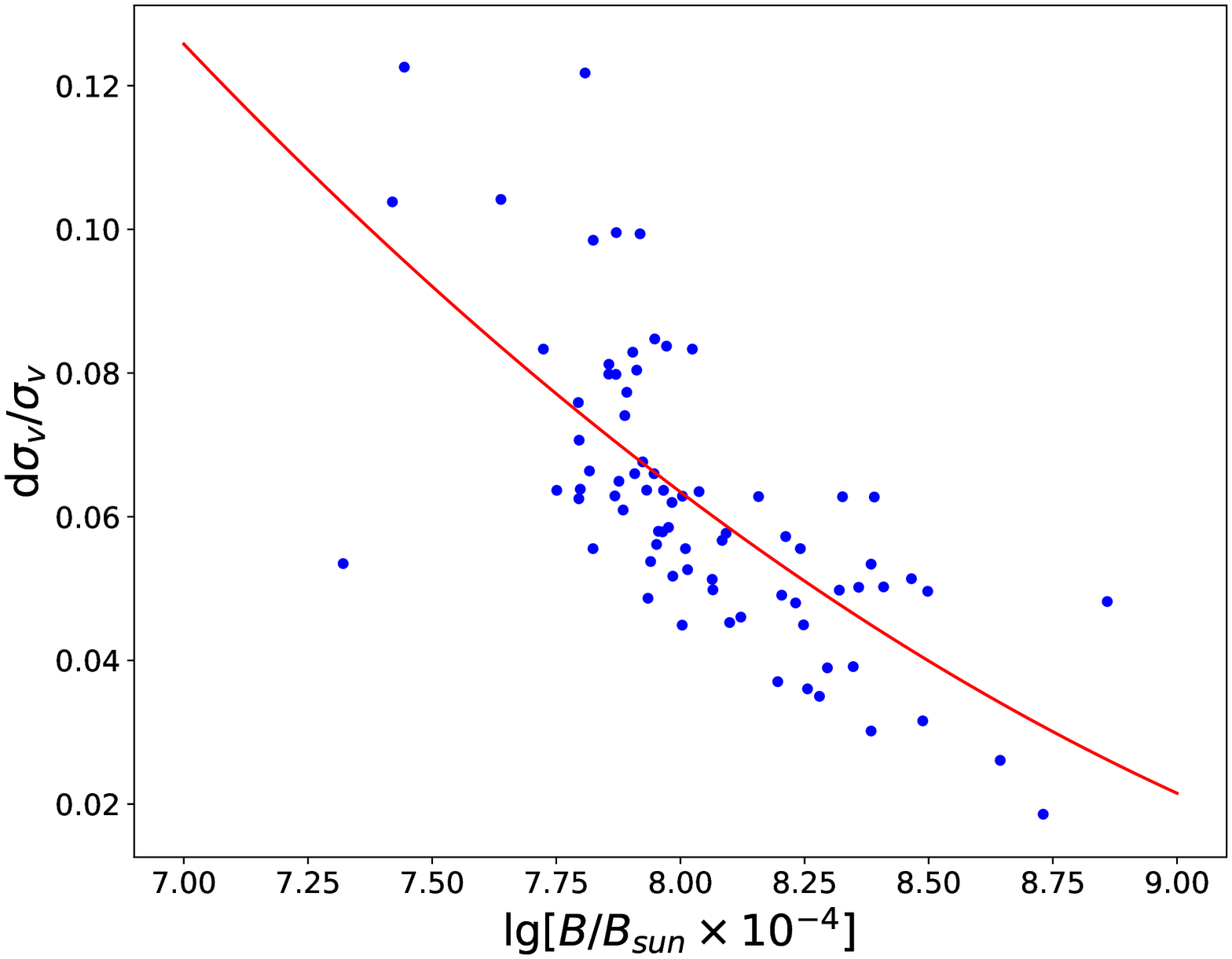}\includegraphics[width=0.4\linewidth]{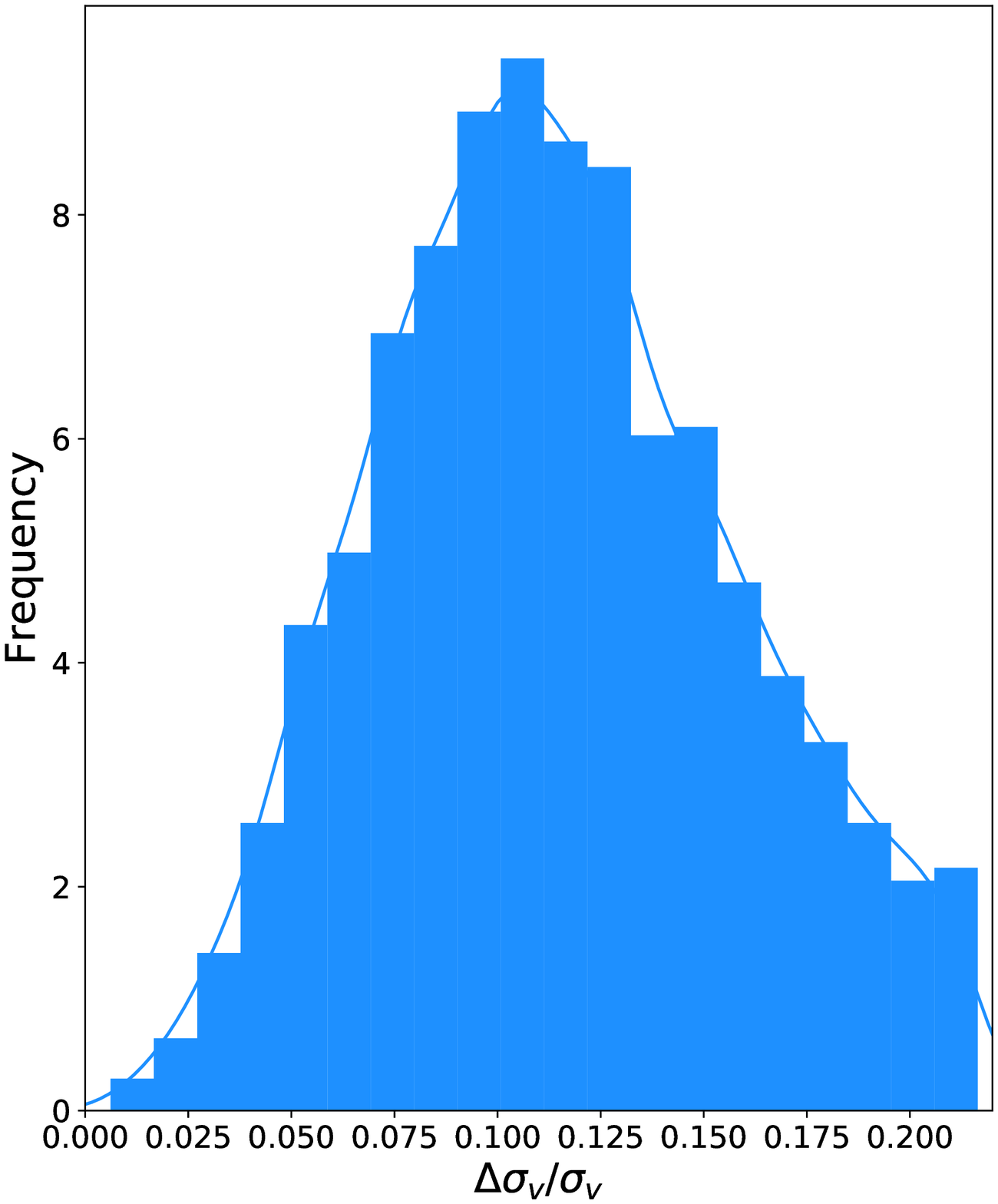}
	\end{center}
	\caption{Left panel: Fractional uncertainty of the velocity
		dispersion ($\Delta\sigma_v/\sigma_v$) as a function of the lens
		surface brightness ($B$) for the SLACS sample, with the best-fitted
		correlation function denoted as the red solid line. Right panel: The
		distribution of the velocity dispersion uncertainty for the
		simulated sample with 5000 SGL systems.}
\end{figure*}

Motivated by \citet{Koopmans06b,Humphrey2010} that indicates early-type galaxies are well characterized by power-law mass distributions in regions covered by the X-ray and lensing observations, we modeled the lensing galaxies with a power-law mass distribution ($\rho \sim r^{-\gamma}$), which leads to
\begin{equation}
\theta_E = 4 \pi \frac{\sigma_{ap}^2}{c^2} \frac{D_{ls}}{D_s} \left(
\frac{\theta_E}{\theta_{ap}} \right)^{2-\gamma} f(\gamma),
\end{equation}
where
\begin{eqnarray}
f(\gamma) &=& - \frac{1}{\sqrt{\pi}} \frac{(5-2 \gamma)(1-\gamma)}{3-\gamma} \frac{\Gamma(\gamma - 1)}{\Gamma(\gamma - 3/2)}\nonumber\\
&\times & \left[ \frac{\Gamma(\gamma/2 - 1/2)}{\Gamma(\gamma / 2)} \right]^2.
\end{eqnarray}
Moreover, the prior of the average logarithmic density slope for the total mass density is modeled as $\gamma=2.01 \pm 1.24$, which is derived from the analysis of massive field early-type galaxies selected from SLACS survey \citep{Koopmans06a,Koopmans06b}. Now, it is important to emphasize some key considerations in our simulation. In order to test GR with the combination of strong lensing and stellar dynamics, additional information including spectroscopic redshift of lenses and sources ($z_{l}$ and $z_{s}$), the Einstein radius ($\theta_{E,\rm obs}$), and spectroscopic velocity dispersion ($\sigma_{ap}$) are demanded. In view of the substantial cost of the dedicated follow-up observations and subsequent endeavors for a sample containing $10^{5}$ SGL systems, concentrating on a carefully selected subset of LSST lenses is more realistic and proper, as has been proposed in recent discussion of multi-object and single-object spectroscopic follow-up to enhance Dark Energy Science from LSST \citep{Chae2003}.
Therefore, referring to the selection criteria presented in \citep{Liu2019,Liu2020}, the final simulated sample is limited to 5000 elliptical galaxies with velocity dispersion of $200 ~km~s^{-1} < \sigma_{ap} < 300 ~km~s^{-1}$. Following the investigation of intermediate-mass lenses to relieve the possible
disagreement between gravitational mass and dynamical mass for the SIS model \citep{Cao2016}, systems with the Einstein radius $\theta_{E,\rm obs} >0.5 arcsec$, and the i-band magnitude $m_{i}<22$ might be difficult to observe precisely. Fig.~2 presents the scatter plot of the simulated lensing systems, where a fair coverage of lens and source redshifts can be noticed. In our analysis, we also considered
three sub-samples defined by different redshift ranges: $z \leq 0.3$, $0.3 < z < 0.65$, $z \geq 0.65$, to expound any noticeable differences in the constraints displayed by lenses from different redshift bins. It should be pointed out that such an approach guarantees that enough data points are contained in each sub-sample and does not involve any physical aspects of the galaxy distribution in redshift.

As for the uncertainty investigation, we adopted the detailed procedure presented in \citep{Liu2020}. With regard to the observed Einstein radius, 32 SGL systems detected by SL2S show a possible correlation between the fractional uncertainty of the Einstein radius and $\theta_{E,\rm obs}$, which means that the lenses with smaller Einstein radii would have larger uncertainty. In our analysis, the fractional uncertainty of $\theta_{E,\rm obs}$ is taken at the level of 8 percent, 5 percent, and 3 percent, respectively for small Einstein radii lenses ($0.5 arcsec < \theta_{E,\rm obs}< 1.0 arcsec$), intermediate Einstein radii lenses ($1.0 arcsec < \theta_{E,\rm obs}< 1.5 arcsec$), and large Einstein radii lenses ($\theta_{E,\rm obs}> 1.5 arcsec$). It is worth noting that the fractional uncertainty of the Einstein radius may reach a level of 3 percent when all of our simulated LSST lenses will be observed with HST-like image quality \citep{Hilbert2009}. For the uncertainty of velocity dispersion, 70 SGL systems from SLACS survey \citep{Bolton08} have been utilized to quantify the correlation between the lens surface brightness in the i band and fractional uncertainty of the velocity dispersion $\Delta \sigma_{v}/\sigma_{v}$. This was an appropriate sample to represent the observations that the future LSST survey might yield.  From Fig.~3, one can see clearly strong evidence of anticorrelation between these two quantities. Then, we take advantage of the best-fitted correlation function derived from the 70 SGL systems to estimate the uncertainty of velocity dispersion for the discoverable lenses in future LSST survey, whose distribution is showed in Fig.~3.

\subsection{Distance from type Ia Supernova observations}

\begin{figure*}
	\begin{center}
		\includegraphics[width=0.55\linewidth]{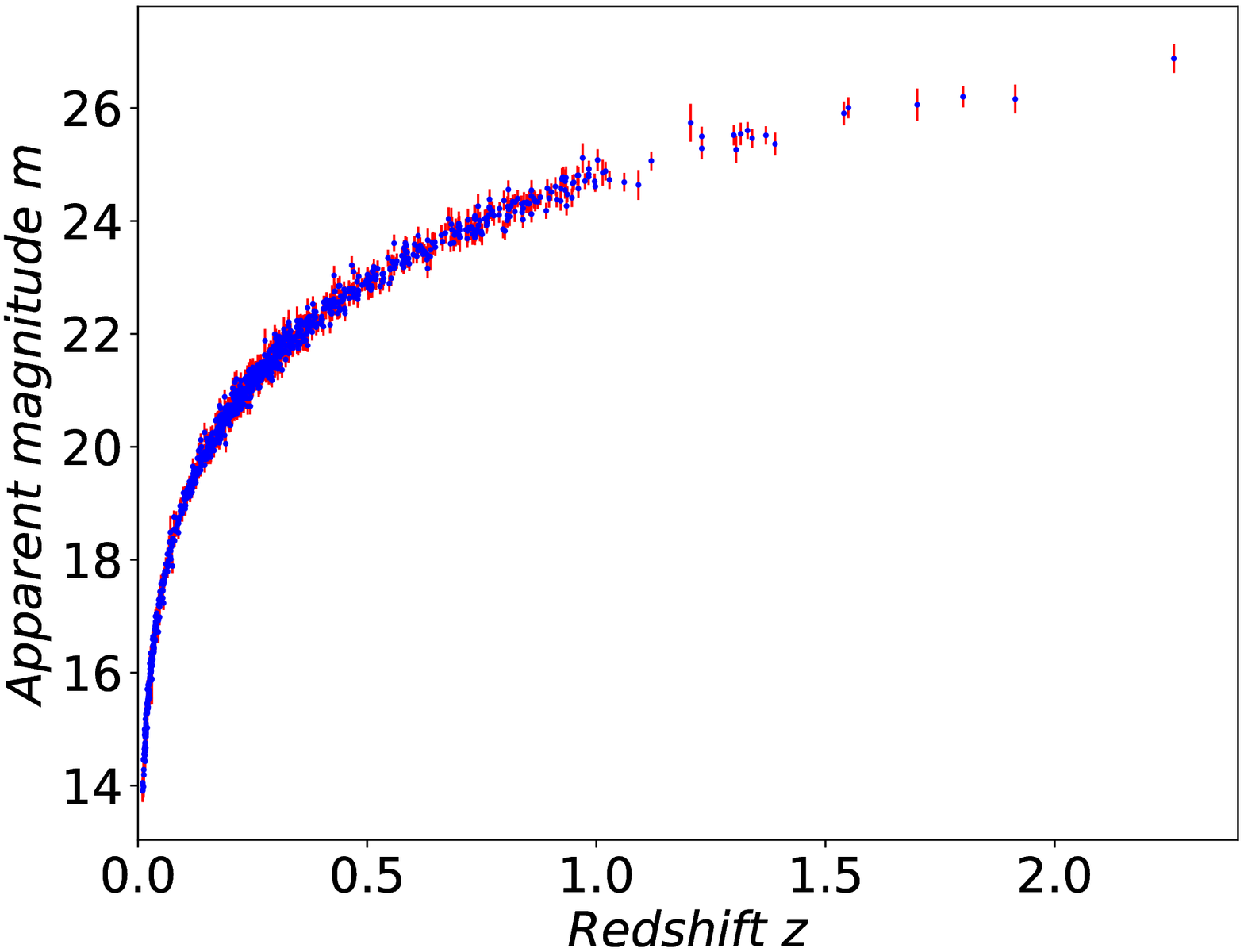}\includegraphics[width=0.4\linewidth]{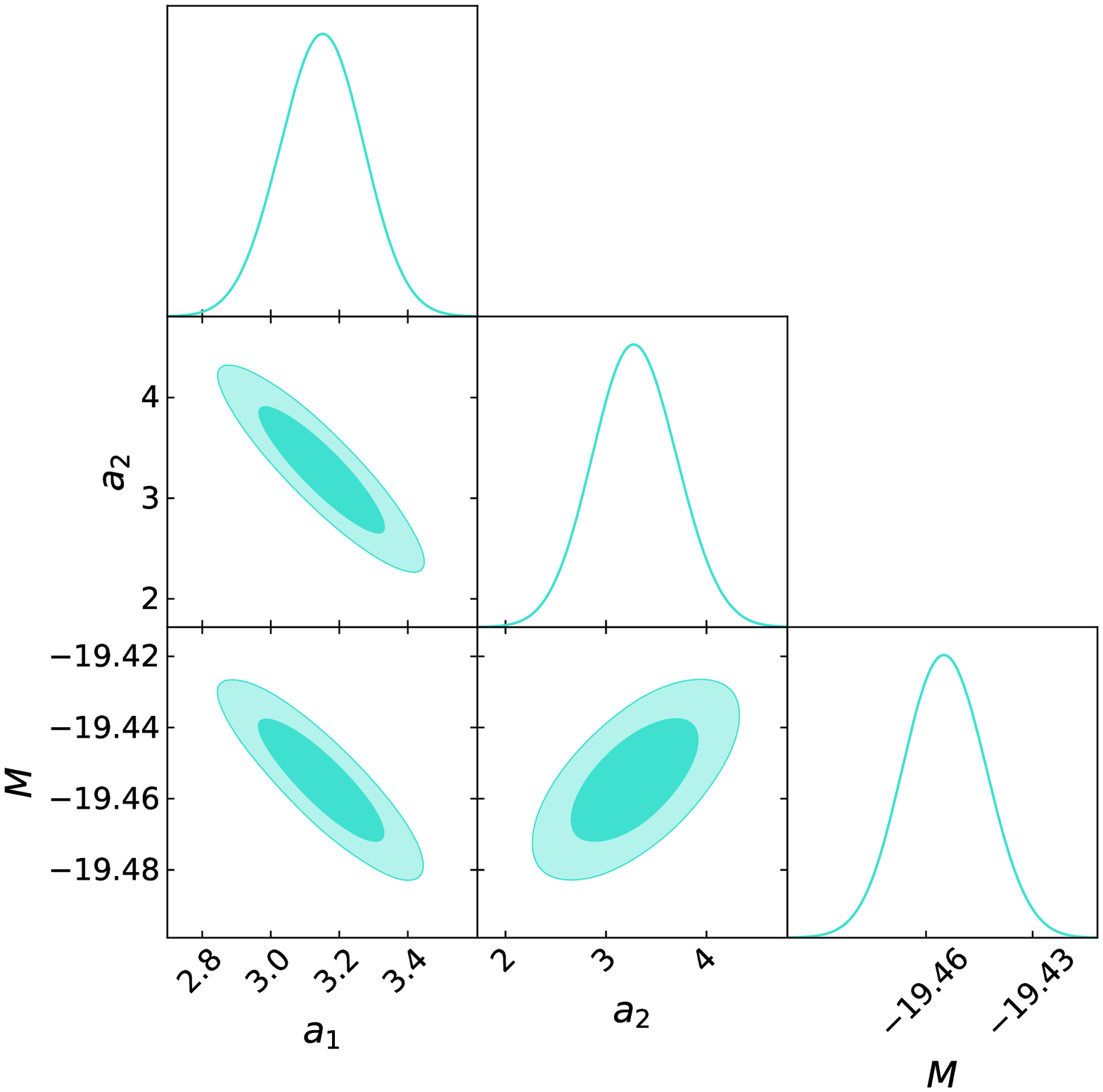}
	\end{center}
	\caption{Left panel: The scatter plot of the 1048 SNe Ia pantheon sample. Observed apparent $B$-band magnitude is plotted against the redshift. Right panel: The 1D probability distributions and 2D contours with 1$\sigma$ and 2$\sigma$ confidence levels for the parameters $a_1$, $a_2$, and M, obtained from 1048 SNe Ia pantheon sample.}
\end{figure*}

In this paper, we assumed a flat concordance $\Lambda CDM$ model to obtain the angular diameter distances $D_{L}$, $D_{S}$, and $D_{LS}$ as customary \citep{Schwab2010,Cao2017}. This was necessary to put constraints on the PPN parameter $\gamma_{PN}$. Theoretically, however, $\Lambda CDM$ is derived within the framework of GR. Therefore, it is important to come up with a model-independent approach, such as the distance reconstruction by standard candles, to derive these angular diameter distances and examine if the constraints would differ from that achieved by the fiducial cosmological model. Meanwhile, considering most current cosmological probes have a limited redshift coverage $z<2.5$, it is necessary to contemplate the possible influence on constraints due to the absence of high redshift samples. Thereupon, besides deriving the cosmological distance for the simulated SGL systems by $\Lambda CDM$ model, we will also adopt a model-independent method to achieve the angular diameter distances, through SN Ia Pantheon data set released by the Pan-STARRS1 (PS1) Medium Deep Survey \citep{Scolnic2018}, which allows us to perform an empirical fit to the luminosity distance measurements. Next, we will introduce the details of reconstruction of the luminosity distance $D_{L}(z)$ from SN Ia.

Containing 1048 SNe Ia measurements over a redshift range $0.01<z<2.3$, the Pantheon catalogue has extended the Hubble diagram to $z=2.26$ compared with high-z data from the SCP survey \citep{Suzuki2012}, the GOODS \citep{Riess2007} as well as CANDELS/CLASH surveys \citep{Rodney2014}. This enables us to calibrate the SGL systems with $z_{S}<2.3$ in the forecasted LSST lenses by SN Ia. Thus, we generated 2619 simulated SGL systems suitable for this procedure. The observed distance modulus of each SN Ia is given by
\begin{equation}
\mu = m_{B}-M_{B}+\alpha \cdot X_{1}-\beta ^{\ast }\cdot C + \Delta M +\Delta B,
\end{equation}
where $m_{B}$ denotes the apparent $B$-band magnitude, $M_{B}$ is the absolute $B$-band magnitude, C represents the color parameter describing the relation between luminosity and color, $X_{1}$ is the light-curve shape parameter quantifying the relation between luminosity and stretch, as well as distance correlations based on the host-galaxy mass ($\Delta M$) and predicted biases from simulations ($\Delta B$) are also considered. According to a new method known as BEAMS with bias corrections \citep{Kessler2017}, the nuisance parameters in Eq.~(27) could be calibrated to zero. Then the observed distance modulus is simply reduced to
\begin{equation}
\mu_{obs} = m_{B}-M_{B}.
\end{equation}
The theoretical distance modulus can be expressed as
\begin{equation}
\mu _{th}=5 \log_{10}(D_{L}/ \rm Mpc)+25.
\end{equation}
As can be seen from Fig.~4, we have presented the dependence of apparent $B$-band magnitude on redshifts, drawn from 1048 SNe Ia observations. In order to achieve the angular diameter distance for SGL systems with $z_{S}<2.3$, we carry out an empirical fit to the luminosity distance measurements, modeled as a third-order logarithmic polynomial expression in $\rm log(1+z)$ \citep{Kessler2017},
\begin{equation}
D_{L}(z)=\ln(10)c/H_{0}(x+a_{1}x^{2}+a_{2}x^{3}),
\end{equation}
where $x=\log(1+z)$, $a_{1}$ and $a_{2}$ are two constant parameters to be optimized and determined by apparent $B$-band magnitude of SNe Ia data. The Markov Chain Monte Carlo (MCMC) method implemented in \emph{emcee} package \footnote{https://pypi.python.org/pypi/emcee} \citep{Foreman-Mackey2013} written in Python 3.7 was used to get the best-fit values and $1\sigma$ uncertainties of the parameters $a_{1}$ and $a_{2}$. From the Hubble diagram of SNe Ia, one can derive the parameters $M_{B}$, $a_{1}$ and $a_{2}$ by minimizing the $\chi ^{2}$ objective function:
\begin{equation}
\chi_{SNe}^{2}=\sum_{i=1}^{1048}\frac{\left [ m_{i}^{obs} -m_{i}^{th}(M_B, a_1, a_2)  \right ]^{2}}{\sigma _{SNe}^{2}},
\end{equation}
where $\sigma _{SNe}$ is the error in SNe Ia observations propagated from the covariance matrix \citep{Scolnic2018}. The 1D probability distributions of each parameters and the 2D contours are presented in Fig.~4.

\begin{table*}
	\begin{center}
		\setlength{\tabcolsep}{1mm}{
		    \begin{threeparttable}
			\begin{tabular}{|c|c|c|c|c|c|c|c|c|}
				\hline
				model & data & $\Lambda \; (kpc)$ & $\gamma_{PN}$ & Galaxy structure parameters  \\
				\hline
				$\gamma=\delta$ & Current mass-selected SGL  & $99.56_{-57.55}^{+97.70}$ & $0.378_{-0.269}^{+0.522}$ &  $\gamma_{0}=2.044_{-0.038}^{+0.045}$, $\gamma_{1}=-0.007_{-0.028}^{+0.021}$  \\
				
				& Forecasted SGL (Full sample) & $211.30_{-95.28}^{+63.44}$ & $0.998_{-0.007}^{+0.003}$ & $\gamma_{0}=2.007_{-0.014}^{+0.010}$, $\gamma_{1}=0.014_{-0.028}^{+0.033}$  \\
				
				& Sub-sample (Logarithmic polynomial) & $213.08_{-90.48}^{+61.59}$ & $0.981_{-0.020}^{+0.021}$ & $\gamma_{0}=2.018_{-0.013}^{+0.013}$, $\gamma_{1}=0.014_{-0.025}^{+0.038}$  \\
				
				& Sub-sample $(z \leq 0.3)$ & $210.04_{-90.90}^{+62.49}$ & $0.998_{-0.025}^{+0.005}$ & $\gamma_{0}=2.003_{-0.033}^{+0.020}$, $\gamma_{1}=0.043_{-0.087}^{+0.104}$  \\
				
				& Sub-sample $( 0.3 < z < 0.65)$ & $212.31_{-90.53}^{+62.10}$ & $0.999_{-0.013}^{+0.007}$ & $\gamma_{0}=2.005_{-0.039}^{+0.047}$, $\gamma_{1}=0.024_{-0.078}^{+0.103}$  \\
				
				& Sub-sample $(z \geq 0.65)$ & $213.21_{-89.65}^{+62.26}$ & $0.999_{-0.097}^{+0.054}$ & $\gamma_{0}=2.021_{-0.141}^{+0.142}$, $\gamma_{1}=0.043_{-0.153}^{+0.195}$  \\
				
				\hline
				$\gamma \neq \delta$ & Current mass-selected SGL & $142.92_{-106.94}^{+97.51}$ & $0.937_{-0.767}^{+1.384}$ &  $\gamma_{0}=2.008_{-0.068}^{+0.069}$, $\gamma_{1}=-0.005_{-0.052}^{+0.045}$, $\delta=2.220_{-0.165}^{+0.168}$     \\
				
				& Forecasted SGL (Full sample) & $213.62_{-86.84}^{+61.57}$ & $0.973_{-0.071}^{+0.027}$ & $\gamma_{0}=2.008_{-0.019}^{+0.012}$, $\gamma_{1}=0.007_{-0.054}^{+0.033}$, $\delta=2.148_{-0.112}^{+0.107}$   \\
				
				& Sub-sample (Logarithmic polynomial) & $203.18_{-88.42}^{+68.17}$ & $0.935_{-0.131}^{+0.051}$ & $\gamma_{0}=2.012_{-0.025}^{+0.018}$, $\gamma_{1}=-0.014_{-0.091}^{+0.038}$, $\delta=2.186_{-0.123}^{+0.100}$   \\
				
				& Sub-sample $(z \leq 0.3)$ & $205.69_{-92.56}^{+66.79}$ & $0.799_{-0.396}^{+0.171}$ & $\gamma_{0}=2.014_{-0.035}^{+0.022}$, $\gamma_{1}=0.052_{-0.203}^{+0.091}$, $\delta=2.156_{-0.120}^{+0.118}$  \\
				
				& Sub-sample $( 0.3 < z < 0.65)$ & $177.25_{-92.24}^{+83.75}$ & $0.547_{-0.369}^{+0.346}$ & $\gamma_{0}=2.045_{-0.057}^{+0.032}$, $\gamma_{1}=0.036_{-0.066}^{+0.046}$, $\delta=2.137_{-0.174}^{+0.158}$  \\
				
				& Sub-sample $(z \geq 0.65)$ & $163.49_{-100.93}^{+93.63}$ & $0.546_{-0.375}^{+0.394}$ & $\gamma_{0}=2.042_{-0.131}^{+0.103}$, $\gamma_{1}=-0.010_{-0.124}^{+0.109}$, $\delta=2.159_{-0.253}^{+0.159}$  \\
				
				\hline
				$\gamma \neq \delta$ & Forecasted SGL (Full sample)\tnote{1} & $212.63_{-92.68}^{+61.70}$ & $0.862_{-0.139}^{+0.115}$ &  $\gamma_{0}=2.024_{-0.074}^{+0.082}$, $\gamma_{1}=-0.005_{-0.018}^{+0.025}$, $\delta=1.938_{-0.149}^{+0.133}$     \\
				
				& Sub-sample (Logarithmic polynomial)\tnote{1} & $205.49_{-88.36}^{+67.64}$ & $0.791_{-0.205}^{+0.157}$ & $\gamma_{0}=2.033_{-0.010}^{+0.011}$, $\gamma_{1}=0.007_{-0.020}^{+0.036}$, $\delta=1.909_{-0.171}^{+0.150}$   \\
				\hline
		    \end{tabular}
		    \begin{tablenotes}
		    \footnotesize
		    \item[1] These two forecasted SGL samples are re-simulated when the gravitational slip with screening effects are considered in the fiducial model.
		    \end{tablenotes}
		    \end{threeparttable}}
		\caption{Summary of the best-fit values with their 1$ \sigma$ uncertainties concerning the screening radius $\Lambda$, PPN parameter $\gamma_{PN}$, and the galaxy structure parameters ($\gamma_{0},\gamma_{1},\delta$) in two different lens models. The results are obtained from the current 99 intermediate-mass lenses, 5000 forecasted LSST lenses, the sub-sample using the logarithmic polynomial cosmographic reconstruction through SNe Ia, as well as three sub-samples covering three different redshift ranges.}
	\end{center}
\end{table*}

\section{Results and discussion}

In this paper, we used a mass-selected sample including 99 SGL systems, 5000 simulated SGL systems expected from future LSST survey and four sub-samples extracted from the forecasted samples with
different selection criteria to probe the gravitational slip with screening effects and the parameters $\gamma_{0},\gamma_{1},\delta$ characterizing the structure of elliptical galaxies. In order to determine these parameters, which are assumed to be the same for all lensing galaxies, we used MCMC method to
sample their probability density distributions based on the likelihood ${\cal L} \sim \exp{(- \chi^2 / 2)}$, where
\begin{equation}
\chi^2 = \sum_{i=1}^{N} \left( \frac{ \bar
	{\sigma}_{*,i}(z_{l,i},z_{s,i},\theta_{E,i}, \theta_{ap,i},
	\sigma_{atm}; \Lambda,\gamma_{PN},\gamma,\delta) - {\sigma}_{ap,i}}{\Delta \bar
	{\sigma}_{*,i}} \right)^2
\end{equation}
was derived from the measured values of velocity dispersion $\sigma_{ap}$, and theoretical prediction Eq.~(\ref{Schwab}) using the observed Einstein radius $\theta_{E,obs}$, and aperture radius $\theta_{ap}$.
For the term $\sigma_{atm}$, which requires the information of seeing recorded by the spectroscopic guide cameras during observing sessions, we have used the seeing summarized in \citet{Cao2016} for the current intermediate-mass sample of SGL systems. For the simulated lensing systems from LSST, the median seeing in i-band, which is 0.75 arcsec reported in \citet{Collett2015} was used. This value was also adopted to calculate the likely yield of observable gravitationally lensed quasars and supernovae based on the properties of LSST \citet{Oguri2010}. The uncertainties of $\sigma_{ap}$ and $\theta_{E,\rm obs}$ were propagated to the final uncertainty ${\Delta \bar {\sigma}_{*,i}}$. Following the SLACS team \citet{Bolton08} and \citet{Cao2016}, the fractional uncertainty of $\theta_{E,\rm obs}$ is taken as 5\%, and the specific strategy for estimating the uncertainty of $\sigma_{ap}$ and $\theta_{E,\rm obs}$ for the simulated LSST lenses can be found in section 3.1. As for the sub-sample of forecasted LSST lenses using the logarithmic polynomial cosmographic reconstruction through SNe Ia, the uncertainties of $D_{L}(z)$  reconstructed from SNe Ia are also propagated to the final uncertainty ${\Delta \bar {\sigma}_{*,i}}$. The numerical results for $(\Lambda,\gamma_{PN})$ and the lens model parameters with 68.3\% confidence level are summarized in Table~1, and the 1D probability distributions and 2D contours with $1\sigma$ and $2\sigma$ confidence levels are displayed in Figs.~5-9.

\subsection{The case with $\gamma = \delta$}

\begin{figure}
	\includegraphics[width=\columnwidth]{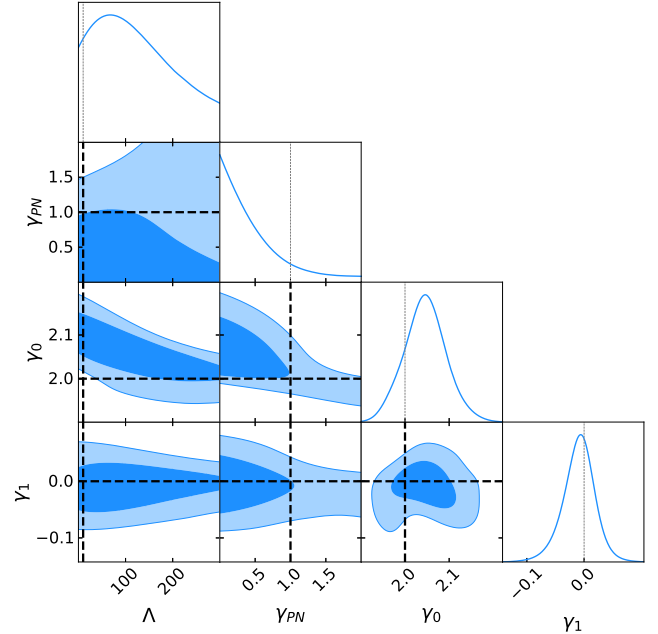}
	\caption{The 1D probability distributions and 2D contours with
		$1\sigma$ and $2\sigma$ confidence levels for the screening radius $\Lambda$,
		the PPN parameter $\gamma_{PN}$, as well as the total-mass density parameters $\gamma_{0}$ and $\gamma_{1}$, obtained from the current sample of 99 intermediate-mass strong gravitational lenses. The black dashed line represents the minimal screening radius at a typical Einstein radius value, GR, and SIS model ($\Lambda=10 \, kpc$, $\gamma_{PN}=1$, $\gamma_{0}=2$, and $\gamma_{1}=0$).}
\end{figure}

\begin{figure*}
	\begin{center}
		\includegraphics[width=\columnwidth]{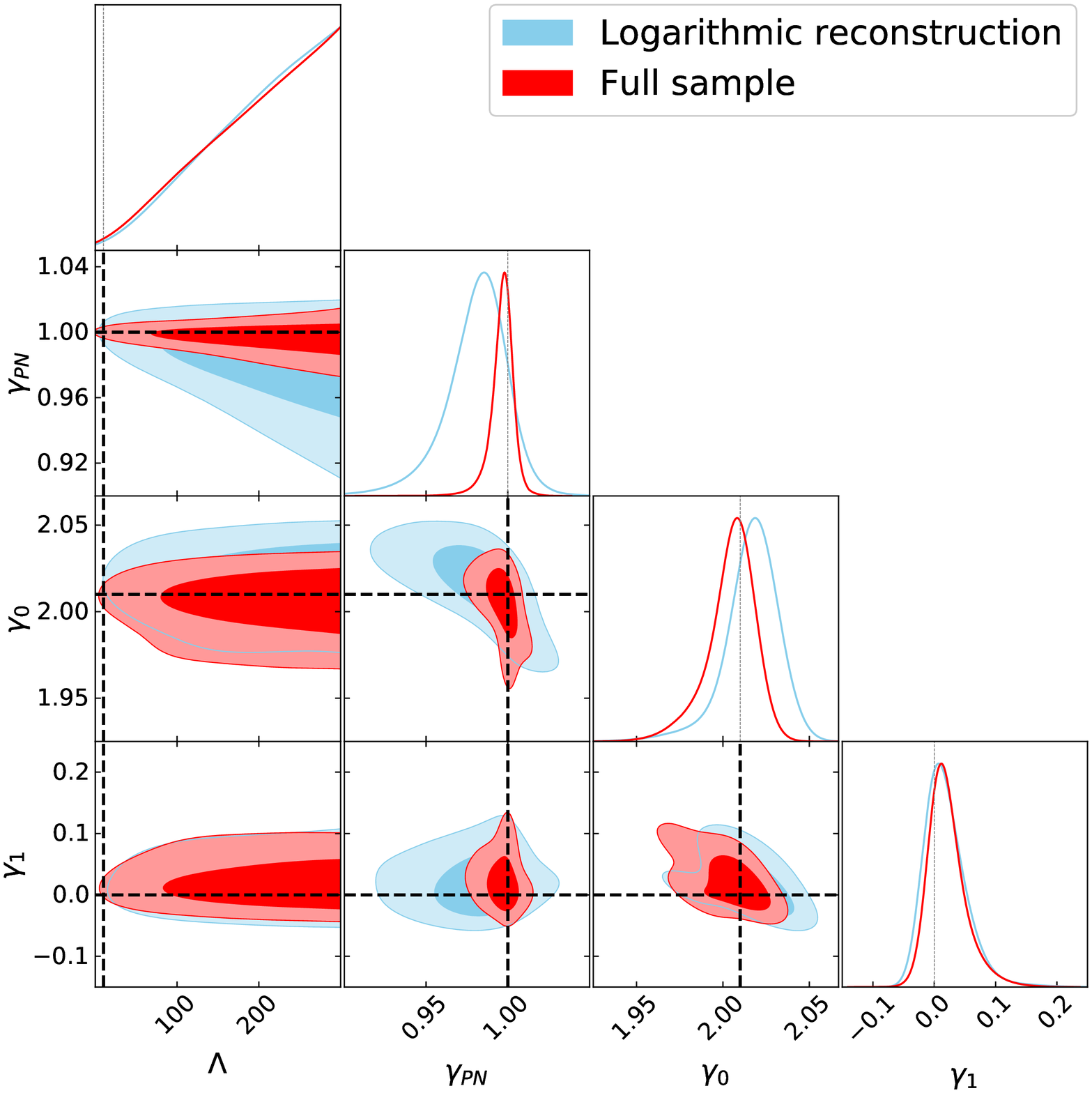}\includegraphics[width=\columnwidth]{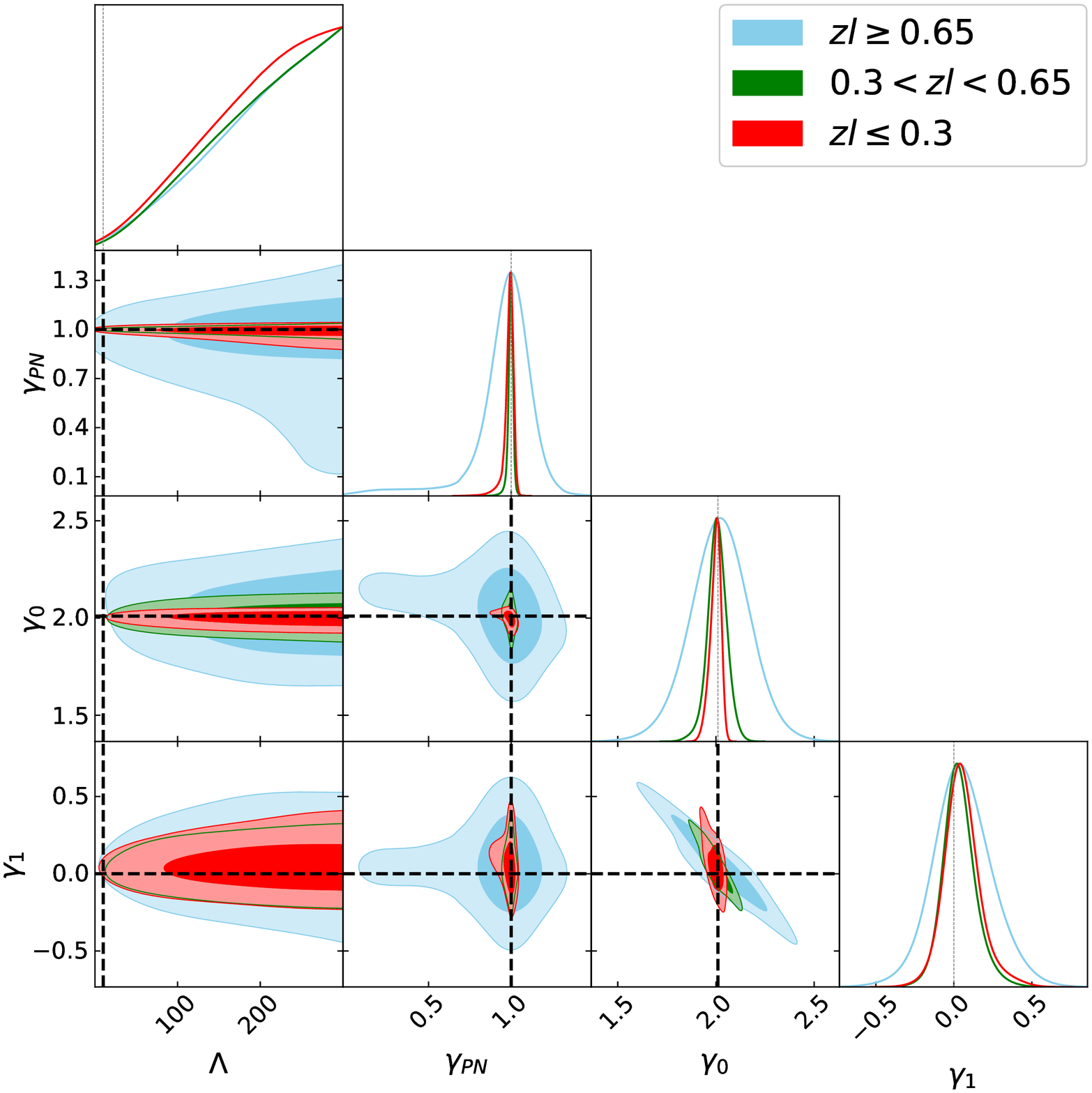}
	\end{center}
	\caption{The 1D probability distributions and 2D contours with
		$1\sigma$ and $2\sigma$ confidence levels for the screening radius $\Lambda$,
		the PPN parameter $\gamma_{PN}$, as well as the total-mass density parameters $\gamma_{0}$ and $\gamma_{1}$. The black dashed line represents the minimal screening radius at a typical Einstein radius value ($\Lambda=10 \, kpc$), GR, and the $\gamma$ prior used in the LSST simulation ($\gamma_{PN}=1$, $\gamma_{0}=2.01$, and $\gamma_{1}=0$). Left panel: The constraints from the full 5000 simulated samples and logarithmic polynomial cosmographic reconstruction. Right panel: The constraints from the sub-samples of SGL systems by three different redshift bins.}
\end{figure*}

In the first case, the luminosity density profile $\nu$ is assumed to follow the total-mass density profile $\rho$, i.e., $\gamma=\delta$. Then, this common density slope is allowed to evolve with redshift:
\begin{equation}
\gamma(z) = \gamma_{0}+\gamma_{1}z,
\end{equation}
where $\gamma_{0}$ is the current value and $\gamma_{1}$ represents the evolution of $\gamma$ with redshift. We assumed that the stellar velocity anisotropy vanishes $\beta=0$ in order to facilitate comparison with previous results. As can be seen from Fig.~5 and Table~1, the current intermediate-mass lenses do not provide stringent constraints on the PPN parameter $\gamma_{PN}$, which will be improved significantly with the simulated full sample of SGL systems. The best-fitted value of $\gamma_{PN}$ obtained from the current sample of 99 lenses is $0.378_{-0.269}^{+0.522}$ with $1\sigma$ confidence level and is still consistent with $\gamma_{PN}=1$ within $2\sigma$. Fig.~5 shows that there is no evident degeneracy between $\gamma_{PN}$ and the screening scale $\Lambda$, but the deviation from GR is more obvious within $1\sigma$ range with the increasing of $\Lambda$. Furthermore, the best-fitted value of the screening radius is $\Lambda=99.56_{-57.55}^{+97.70} \; kpc$. This suggests that there may exist some characteristic scale for the current mass-selected sample, beyond which the modification of GR is possible. It is worth noting that a recent study \citep{Liu2022} got the constraint $\gamma_{PN}$ ($1.455_{-0.127}^{+0.154}$) with the same parametrized models of $\gamma$, where the GR is excluded above $2\sigma$ range. The constraints on the parameters $\gamma_{0}$ and $\gamma_{1}$ are tight: $\gamma_{0}=2.044_{-0.038}^{+0.045}$, $\gamma_{1}=-0.007_{-0.028}^{+0.021}$, which is consistent with the singular isothermal sphere model ($\gamma_{0}=2,\gamma_{1}=0$) within $1\sigma$ range and is also in agreement with similar results, obtained by the others from a mass-selected sample of 80 intermediate-mass lenses ($\gamma_{0}=2.115 \pm 0.072$, $\gamma_{1}=-0.091 \pm 0.154$; \citet{Cao2016}), Union2.1+Gamma ray burst(GRB)+SGL ($\gamma_{0}=2.04_{-0.06}^{+0.08}$, $\gamma_{1}=-0.085_{-0.18}^{+0.21}$; \citet{Holanda2017}) and BAO+SGL ($\gamma_{0}=2.094_{-0.056}^{+0.053}$, $\gamma_{1}=-0.053_{-0.102}^{+0.103}$; \citet{Li2016}). In addition, our results indicate that the total density profile of the current early-type galaxies with intermediate velocity dispersions ($200 ~km~s^{-1} < \sigma_{ap} < 300 ~km~s^{-1}$) have showed no significant evolution over the cosmic time (at least up to $z \sim 1$).

On the other hand, as can be seen from Fig.~6 and Table~1, the simulated full sample provides more stringent constraints on $\gamma_{PN}$ ($0.998_{-0.007}^{+0.003}$) with 0.5\% precision, which is in perfect agreement with $\gamma_{PN}=1$ assumed in simulations. One can see from Figs. 6-7 that much more stringent constraints on $\gamma_{PN}$ would be achieved using the strong lensing systems detectable in the future surveys. For comparison, our results are similar to the results $\gamma_{PN}=1.000_{-0.0023}^{+0.0025}$ obtained in  \citet{Cao2017} with 53000 simulated SGL systems 
meeting the redshift criteria $0<z_{l}<z_{s}\leq 1.414$, where the galaxy structure parameters have been fixed: $\gamma=\delta=2, \beta=0$. The constraint from the sub-sample combined with SNe Ia calibrated as standard candles is $\gamma_{PN} = 0.981_{-0.020}^{+0.021}$ with 2\% precision. However, it is noticeable that the central value of $\gamma_{PN}$ deviates a bit more from $\gamma_{PN}=1$, in comparison to the constraint obtained with the full sample of simulated SGL systems, which may signal some systematics present. In Table~1 and Fig.~6, we also display the results obtained on three sub-samples with different lens redshift bins. One may see that the values are in full agreement with each other and GR ($\gamma_{PN}=1$) is still included within $1\sigma$ range, which is consistent with the assumption that GR is valid when simulating the LSST SGL systems. Interestingly, the degeneracy between the PPN parameter and the screening scale $\Lambda$ showed in Fig.~6 is quite similar to the results from \citet{Jyoti2019,Abadi2021}, where the authors applied time delay in SGL to acquire the constraint on $\gamma_{PN}$ as a function of the physical cutoff scale $\Lambda$ rather than carrying out full MCMC analysis for the parameters ($\gamma_{PN}, \, \Lambda$, and $\gamma$). Furthermore, the minimal screening radius corresponding to 68\% C.L. reaches 42 kpc and 120 kpc for the current and forecasted SGL samples respectively, which is notably larger than the minimal screening scale at a typical Einstein radius value, $\Lambda \geq R_E \approx 10$ kpc \citep{Jyoti2019,Abadi2021}. Considering that the screening length $\Lambda$ is bigger than the Einstein radius of the lensing galaxy, the advantage of the observed velocity dispersion of the intermediate-mass elliptical galaxies to probe gravitational slip under this screened MG is limited. It is still possible since gravitational lensing probes the projected mass i.e. a cylinder along the line of sight. Up to the scales of $\Lambda \sim 300\; kpc$, we did not find any typical screening radius for the samples from simulated LSST lenses, beyond which MG is relevant. Besides, the constraints on the total density profile of early type galaxies indicate that the measurement precision of the current value is expected to be improved to 0.6\% from the full simulated SGL sample, but the accuracy of $\gamma_1$ does not seem sensitive to the sample size. The SIS model is still included within $1\sigma$ range for all samples derived from the simulated SGL systems, which is in good agreement with the prior $\gamma=2.01 \pm 1.24$ used to generate the forecasted LSST SGL systems.

\begin{figure}
	\includegraphics[width=\columnwidth]{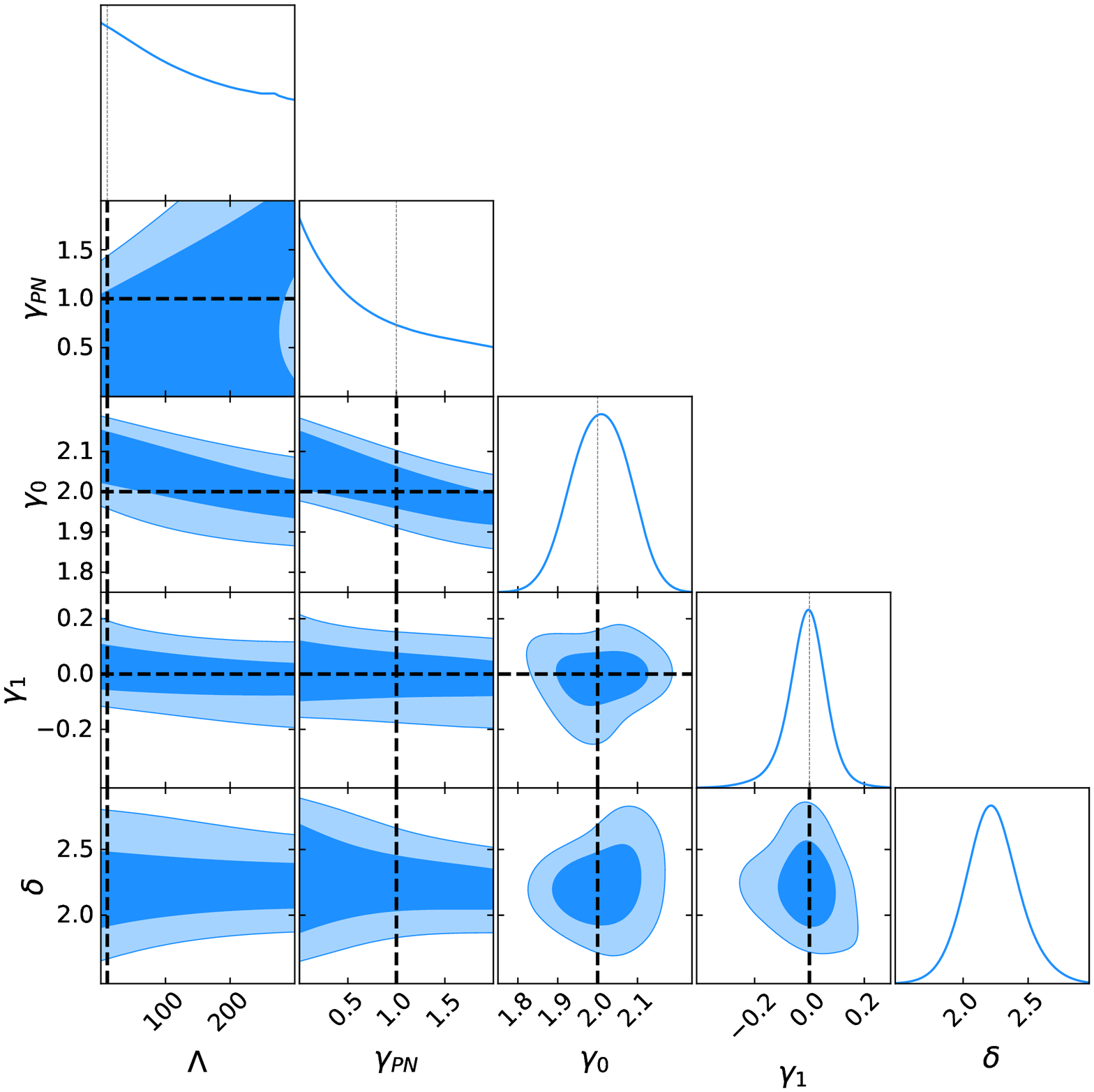}
	\caption{The 1D probability distributions and 2D contours with
		$1\sigma$ and $2\sigma$ confidence levels for the screening radius $\Lambda$, the PPN parameter $\gamma_{PN}$, the total-mass density parameters $\gamma_{0}$ and $\gamma_{1}$, as well as the luminosity density parameter $\delta$, obtained from the current sample of 99 intermediate-mass strong gravitational lenses. The black dashed line indicates the minimal screening radius at a typical Einstein radius value, GR, and SIS model ($\Lambda=10 \, kpc$, $\gamma_{PN}=1$, $\gamma_{0}=2$, and $\gamma_{1}=0$).}
\end{figure}

\subsection{The case with $\delta \neq \gamma$}

\begin{figure*}
	\begin{center}
		\includegraphics[width=\columnwidth]{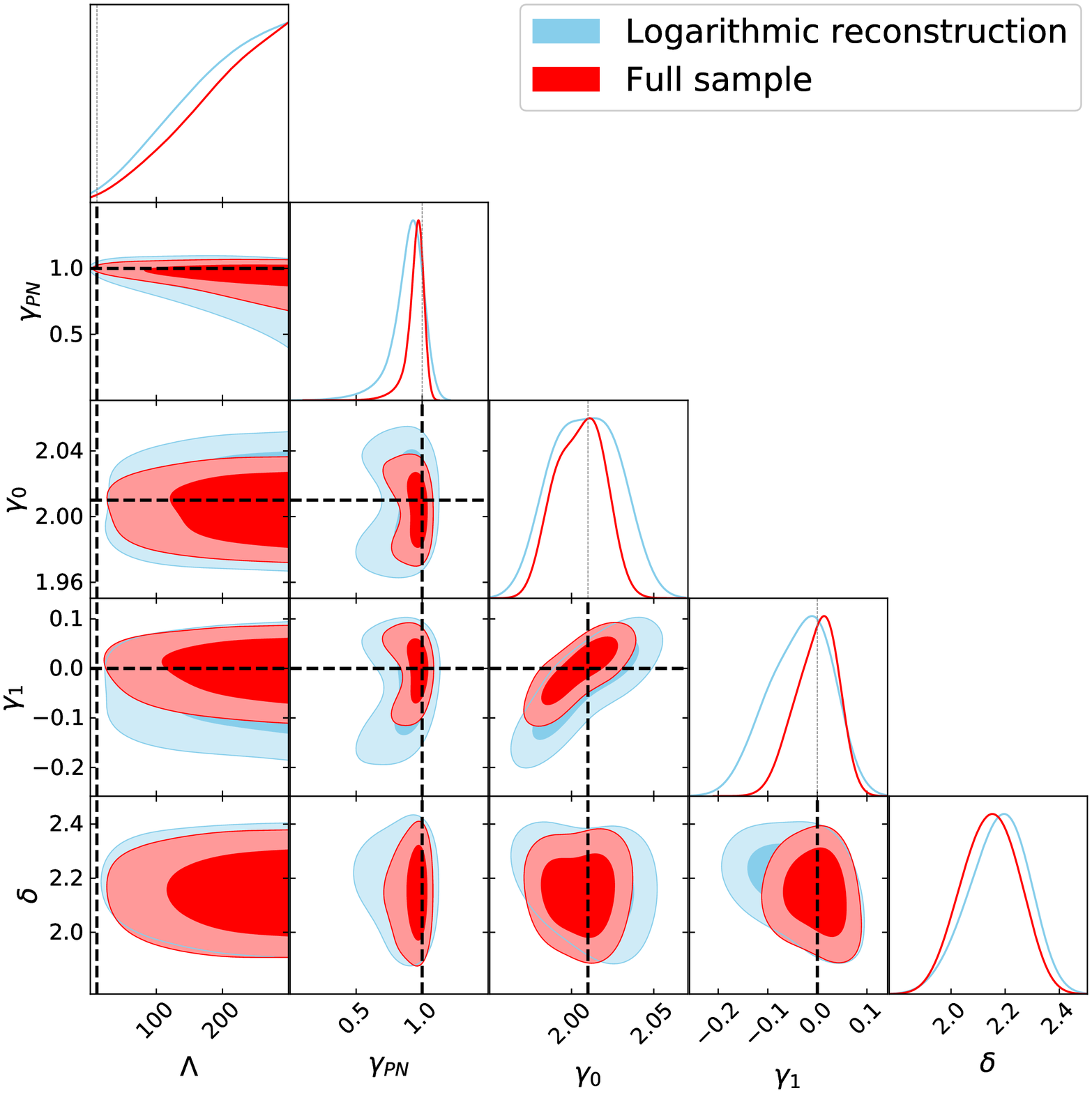}\includegraphics[width=\columnwidth]{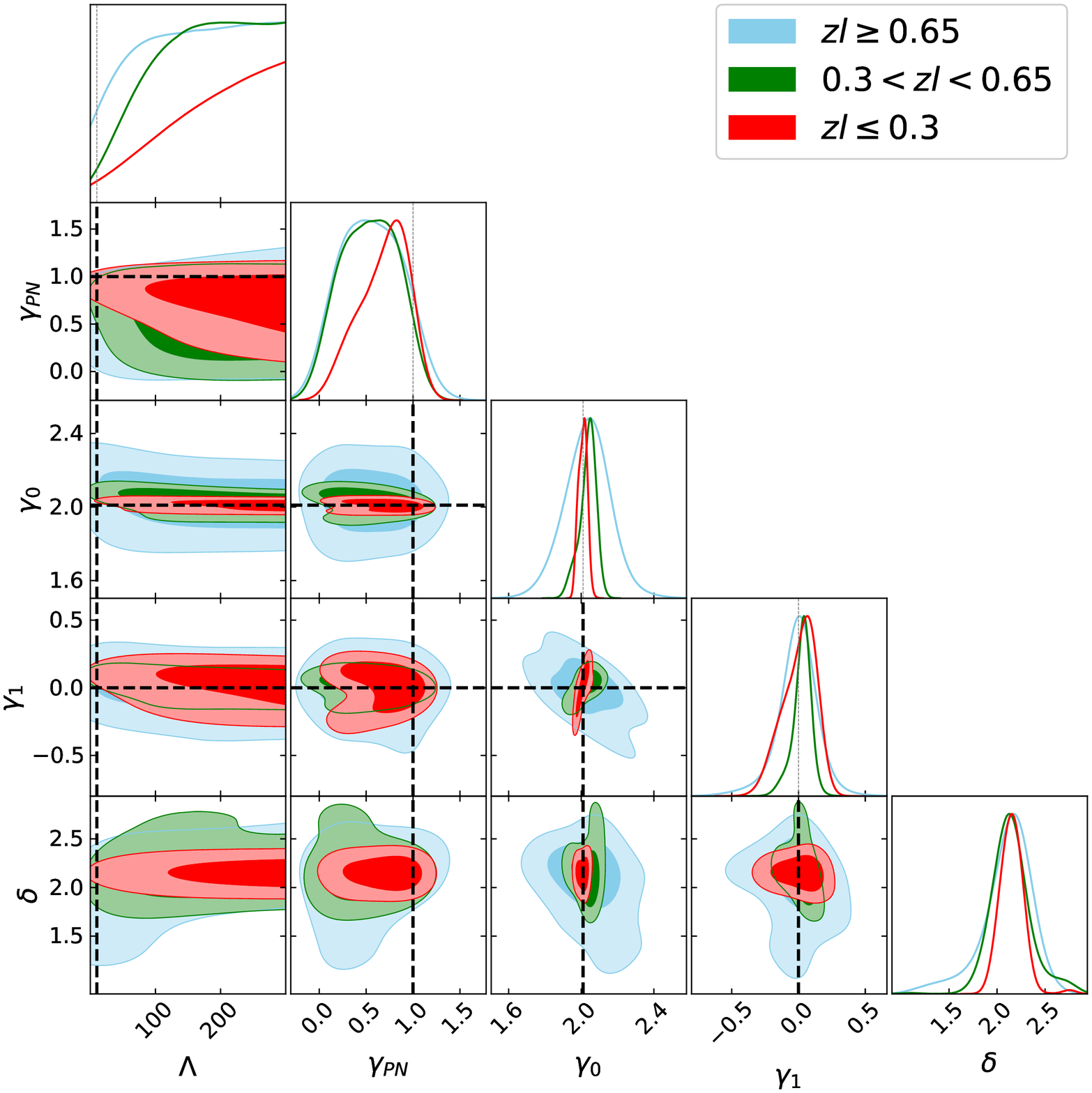}
	\end{center}
	\caption{The 1D probability distributions and 2D contours with $1\sigma$ and $2\sigma$ confidence levels for the screening radius $\Lambda$, the PPN parameter $\gamma_{PN}$, the total-mass density parameters $\gamma_{0}$ and $\gamma_{1}$, as well as the luminosity density parameters $\delta$. The black dashed line represents the minimal screening radius at a typical Einstein radius value ($\Lambda=10 \, kpc$), GR, and the $\gamma$ prior used in the LSST simulation ($\gamma_{PN}=1$, $\gamma_{0}=2.01$, and $\gamma_{1}=0$). Left panel: The constraints from the full 5000 simulated samples and logarithmic polynomial cosmographic reconstruction. Right panel: The constraints from the sub-samples of SGL systems by three different redshift bins.}
\end{figure*}

In the second case, the luminosity density profile is allowed to be different from the total-mass density profile, i.e., $\delta \neq \gamma$, and the stellar velocity anisotropy $\beta$ has been marginalized over a Gaussian distribution, $\beta=0.18 \pm 0.13$, which is also broadly applied in the literature \citep{Gerhard2001,Bolton2006,Schwab2010,Cao2017}. With the same assumption as above, the total density slope was allowed to evolve as a function of redshift, $\gamma(z) = \gamma_{0}+\gamma_{1}z$. All values of the estimated parameters in the screened MG model are displayed in Table~1 and illustrated in Figs.~7-9. For the current 99 intermediate-mass lenses, the constraint on $\gamma_{PN}$ is still very weak: $\gamma_{PN}=0.937_{-0.767}^{+1.384}$, but agrees with GR within $1\sigma$.
In the case of $\delta \neq \gamma$, it seems that there is no obvious degeneracy between $\gamma_{PN}$ and the screening radius $\Lambda$, meanwhile, the constraint on $\Lambda$ displayed in Fig.~7 shows no characteristic scale of screening radius for the present SGL sample, compared with the result inferred in the case of $\delta=\gamma$. Nevertheless, the degeneracy between $\gamma_{PN}$ and $\gamma_0$ (and to a smaller degree with $\gamma_1$) is noticeable in Fig.~7, where a steeper present total mass density profile will contribute to a larger value for the PPN parameter. Performing fits on the current mass-selected sample, the 68\% C.L. uncertainties on the three galaxy structure parameters are $\gamma_{0}=2.008_{-0.068}^{+0.069}$, $\gamma_{1}=-0.005_{-0.052}^{+0.045}$, $\delta=2.220_{-0.165}^{+0.168}$. It is interesting to note that the constraints on the mass-density exponents are consistent with that derived in the case of $\delta=\gamma$ and the singular isothermal sphere is still favoured within $1\sigma$. This indicates that from the perspective of stellar dynamics, it is effectively similar to characterize the mass distribution in the lensing galaxies with intermediate velocity distributions ($200 ~km~s^{-1} < \sigma_{ap} < 300 ~km~s^{-1}$) by both $\delta = \gamma$ and $\delta \neq \gamma$ models. Additionally, we also get a relatively smaller central value of the luminosity density profile, as compared to the results obtained from 53 SLACS lenses ($\delta=2.40 \pm 0.11$; \citep{Schwab2010}), as well as 80 intermediate mass lenses observed by SLACS, BELLS, LSD, and SL2S ($\delta=2.485_{-1.393}^{0.445}$; \citep{Cao2017}). However, in view of no obvious cosmic evolution showed in the total mass density parameter $\gamma$ ($\gamma = \gamma_0+\gamma_1z \approx 2.008$), a model where mass traces light ($\gamma=\delta$) seems to be eliminated at >68\% confidence level and our analysis partly suggests the presence of dark matter in the form of a mass component distributed differently from light.

The full sample of simulated SGL provides more stringent constraints on the PPN parameter $\gamma_{PN}=0.973_{-0.071}^{+0.027}$, lens models $\gamma_{0}=2.008_{-0.019}^{+0.012}$, $\gamma_{1}=0.007_{-0.054}^{+0.033}$, and $\delta=2.148_{-0.112}^{+0.107}$, compared with the results generated from the current sample. As can be seen from Table~1, there is a good consistency between the current mass-selected SGL and the full sample of forecasted SGL. However, the parameter $\gamma_{PN}$ obtained from the cosmographic reconstruction, which is $\gamma_{PN}=0.935_{-0.131}^{+0.051}$, deviates a bit more from $\gamma_{PN}=1$ within $1\sigma$ in comparison to the constraint from the full simulated sample. It is worth noting that this slight deviation from GR is also in agreement with a similar result achieved in the case $\delta = \gamma$ ($\gamma_{PN}=0.981_{-0.020}^{+0.021}$). This illustrates the possibility that using the SNe Ia pantheon sample as a precise distance estimator, through the logarithmic polynomial cosmographic reconstruction, may provide a valuable supplement to the a priori assumed cosmology in probing gravitational slip over the redshift range $0<z<2.5$. From the constraints acquired on three sub-samples showed in Fig.~8 and Table~1, the results on $\gamma_{PN}$ are different from that obtained in the case of $\delta = \gamma$. Namely, one can see that in the samples of the simulated galaxies differing by redshift bin one has a different distribution of the PPN parameter: $\gamma_{PN} =0.799_{-0.396}^{+0.171}$ for $z \leq 0.3$, $\gamma_{PN} = 0.547_{-0.369}^{+0.346}$ for $0.3 < z < 0.65$, and  $\gamma_{PN} = 0.546_{-0.375}^{+0.394}$ for $z \geq 0.65$, which display obvious deviation from $\gamma_{PN}=1$ especially for the case of $0.3 < z < 0.65$ and $z \geq 0.65$ but GR is still valid within $2\sigma$. It is interesting to note that such noticeable impact of two different lens models on $\gamma_{PN}$ is also present in the current SGL sample ($\gamma_{PN}=0.378_{-0.269}^{+0.522}$ vs. $\gamma_{PN}=0.937_{-0.767}^{+1.384}$), which indicates that the constraints on the PPN parameter may be sensitive to the choice of the mass distribution of early-type galaxies. According to the constraints on screening radius presented in Fig.~8 and Table~1, we still do not find any characteristic cutoff scale for all the simulated SGL samples, and the minimal screening radius is lying in the range $\Lambda \in [36,127]\; kpc$ in the case of $\delta \neq \gamma$, which is obviously larger than the minimal screening radius considered in the previous work i.e. $\Lambda \geq R_E \approx 10$ kpc; \citep{Jyoti2019,Abadi2021}. With respect to the mass density parameters $\gamma_{0}$, $\gamma_{1}$, and $\delta$, we can see clearly from Table~1 that the constraints from LSST lenses agree well with that from the current SGL sample within 68.3 percent. Furthermore, the future LSST lenses will improve the measuring precision of the present mass density parameter $\gamma_{0}$ to 0.7\% and the luminosity density parameter $\delta$ to 5\% in comparison to the precision of $\gamma_{0}$ (3.4\%) and $\delta$ (7.5\%) obtained from the current SGL sample.

There are several sources of systematics we do not consider in the 
above analysis. First of all, the validity of GR was assumed in the simulation of forecasted LSST sample. In order to further test the effectiveness of our method, we considered to re-simulate the LSST lenses with modified gravity effects present in the fiducial model. The correction due to the screening effect has been involved in the lensing potential (Eq.~(4)), while the connection between the observed Einstein angle $\theta_{E,\rm obs}$ and the Einstein angle in GR $\theta_{E,\rm GR}$ has been expressed in Eq.~(15). More specifically, in the simulation procedure, the numerical solution of $\theta_{E,\rm obs}$ can be solved through Eq.~(15) and the expression of $\theta_{E,\rm GR}$ is modeled by Eq.~(25)-(26). Furthermore, the priors of the screening scale and the deviation from GR are adopted in the fiducial model as $\Lambda=100 \; kpc$ and $\gamma_{PN}=0.97 \pm 0.09$, which is consistent with $|\gamma_{\rm PN}-1| \leq 0.2 \times (\Lambda/100\,\rm kpc)$ reported in \citep{Collett2018,Jyoti2019}. Based on the simulated 5000 strong lenses with modified gravity effect and extended power-law lens as fiducial models, we obtained the constraints on the parameters $(\Lambda, \gamma_{PN}, \gamma_{0}, \gamma_{1},\delta)$ displayed in Table~1 and Fig.~9. Note that the PN parameter and lens parameters derived from the simulated strong lenses in our analysis, $\gamma_{PN}=0.862_{-0.139}^{+0.115}$, $\gamma_{0}=2.024_{-0.074}^{+0.082}$, $\gamma_{1}=-0.005_{-0.018}^{+0.025}$, and $\delta=1.938_{-0.149}^{+0.133}$ are in good agreement with the above priors on $\gamma_{PN}$ and lens mass density profile at 68.3\% confidence level. The strong degeneracy between $\Lambda$ and $\gamma_{PN}$ could also be seen from Fig.~9. Secondly, in this study two power-law lens models have been adopted to connect the observed velocity dispersion to the gravitational slip under screening effects, which presents the direct test of GR within screening scales of $\Lambda=10 - 300 \; kpc$. As was noted in the previous works \citep{Schwab2010,Chen2019,Cao2017,Liu2020,Wong2020,Liu2022}, our analysis indicates that the lens mass modeling may have an apparent influence on the estimation of cosmological parameters such as the screening scale $\Lambda$ and PN parameter $\gamma_{PN}$. Therefore, besides benefitting from the dramatically increasing number of SGL systems observed by future optical surveys, more appropriate modeling of the lens mass will also contribute to the understanding of lens parameters and reducing the uncertainty of derived cosmological parameters. For instance, although the effectiveness of power-law density profiles has been widely proved in describing the observed early-type galaxies within a few effective radii \citep{Wang2018}, the scatter of other galaxy density parameters could be an important source of systematic errors on the final results. An influential paper by \citet{Navarro1996,Navarro1997} suggested that the Navarro-Frenk-White (NFW) profile can provide a good approximation to the the density profile of dark matter (DM) halos, which has found widespread astrophysical applications in the literature \citep{Bullock2001,Komatsu2011,Koyama2016,Collett2018}. Such spherically averaged density profile is well described by a double power-law relation, which resembles dark matter halo with $r^{-3}$ in the outer regions and $r^{-1}$ at small radii \citep{Cardone2010}. However, the joint strong lensing and dynamical analysis strongly support that the total mass profile is very close to isothermal ($\gamma\sim 2$), although neither the stellar component nor dark matter halo is of a simple power law \citep{Treu2010}. Detailed solutions to the lensing and dynamical properties of lenses (such as the total mass, velocity dispersion and Einstein radius) and the gravitational slip under the NFW profile would be an independent work in our future studies.

\section{Conclusions}

\begin{figure}
	\includegraphics[width=\columnwidth]{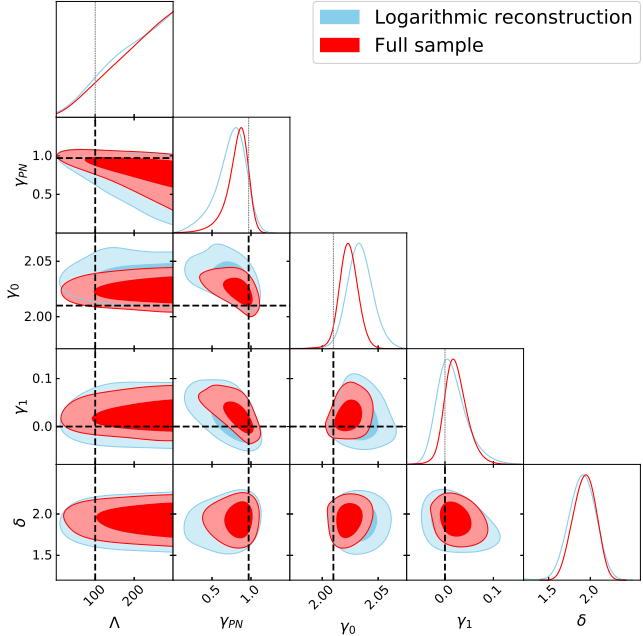}
	\caption{The 1D probability distributions and 2D contours with
		$1\sigma$ and $2\sigma$ confidence levels for the screening radius $\Lambda$, the PPN parameter $\gamma_{PN}$, the total-mass density parameters $\gamma_{0}$ and $\gamma_{1}$, as well as the luminosity density parameter $\delta$, obtained from the re-simulated 5000 LSST lenses with gravitational slip included in the fiducial model. The black dashed line indicates the $\Lambda$, $\gamma_{PN}$, and $\gamma$ priors used in the LSST simulation ($\Lambda=100 \, kpc$, $\gamma_{PN}=0.97$, $\gamma_{0}=2.01$, and $\gamma_{1}=0$).}
\end{figure}

In this work, we have studied the gravitational slip under a phenomenological model of gravitational screening, where GR is maintained for small radii and the departures from GR take the form of a gravitational slip beyond the screening scale $\Lambda$. Based on mass-selected galaxy-scale strong gravitational lenses from the SLACS, BELLS, LSD, as well as SL2S surveys and simulated future measurements of 5000 galaxy-scale SGL from the forthcoming Large Synoptic Survey Telescope (LSST) survey, we were able to evaluate this screened MG model with screening length $\Lambda=10 - 300 \; kpc$, which is broad enough to cover one typical massive galaxy. This is also the first attempt to use the stellar kinematics of SGL systems to assess constraints on the PPN parameter $\gamma_{PN}$, screening radius $\Lambda$ within this screening range under two different lens models simultaneously. Here we summarize our main conclusions in more details.

Considering two different lens models where the total mass density and luminosity density of lensing galaxies are modeled as power-law density profiles $\rho(r) = \rho_0 (r/r_0)^{-\gamma}$ and $\nu(r) = \nu_0 (r/r_0)^{-\delta}$ respectively and $\gamma$ is assumed to evolve with redshift, our results indicate that the current intermediate mass early-type galaxies are not able to provide tight constraints on the PPN parameter in this new theories of modified gravity, but GR ($\gamma_{PN}=1$) is still valid with screening length $\Lambda=10 - 300 \; kpc$ in both lens models. On the other hand, our work studies the complementary range $\Lambda \geq R_E$ compared with previous researches \citep{Bolton2006,Smith2009,Schwab2010,Collett2018,Liu2022} where the screening is assumed to take place within the galaxy. Then, one interesting thing is to figure out if there exists any specific cutoff scale for galaxy-scale SGL systems beyond which MG is relevant. The constraint achieved in the case of $\gamma=\delta$ shows a significant value of $\Lambda\sim 100$ kpc, beyond which the departure from $\gamma_{PN}$ is noticeable within $1\sigma$ confidence level. Nevertheless, in the case of $\gamma \neq \delta$, the current sample shows no specific cutoff value in the range $10 \; kpc <\Lambda<300 \; kpc$. It supports the claim that the intermediate mass lenses may shed new light on testing the validity of GR under this screened MG model. In addition, the 68 \% confidence level constraints on $\Lambda$ and $\gamma_{PN}$ are quite different depending on the lens model. Setting the luminosity profile of elliptical galaxies as a free parameter, we obtained larger best fit values of $\Lambda=142.92_{-106.94}^{+97.51}\; kpc$ and $\gamma_{PN}=0.937_{-0.767}^{+1.384}$ compared with  $\Lambda=99.56_{-57.55}^{+97.70} \;kpc$ and $\gamma_{PN}=0.378_{-0.269}^{+0.522}$ in the case of $\gamma = \delta$. Moreover, in this paper, we assessed the constraints for the total mass-profile and light-profile shapes of elliptical galaxies. Allowing for the cosmic evolution of the total mass density profile exponent in the form of $\gamma =\gamma_0+\gamma_1 z$, there is no obvious evidence suggesting that the total density profile of intermediate mass early-type galaxies has become steeper over cosmic time (up to $z\sim 1$), and the singular isothermal sphere model is well favored by the current mass-selected sample in both lens models.

Furthermore, we elaborated what kind of results one could acquire making use of the future measurements of a well-selected sample containing 5000 LSST lenses. The final results imply that much more severe constraints on the $\gamma_{PN}$ will be achieved with $10^{-2} \sim 10^{-3}$ precision in the regime of screening radii $\Lambda = 10 - 300 \; kpc$. Benefiting from LSST's wide field of view and sensitivity, the SGL systems detectable in the future can be very helpful for testing GR on kpc-Mpc scales in modified theories of gravity. Interestingly, the degeneracy between the screening scale $\Lambda$ and the PPN parameter $\gamma_{PN}$ derived from the full simulated SGL samples and four sub-samples is very similar to the results presented in \citet{Jyoti2019,Abadi2021}, where the authors do not carry out full MCMC analysis in parameter space. Meanwhile, we still do not find any particular cutoff scale for these simulated SGL samples, which is consistent with the assumption that GR is valid and no screening effects are involved in simulating the LSST SGL systems.
With the increasing number of available galaxy-scale lenses, our results imply that it would be advantageous to use velocity dispersion measurements of the intermediate-mass elliptical galaxies to probe departures from GR under the screened MG model, where the gravitational slip is modeled as a step-wise discontinuous phenomenon with the screening radius $\Lambda$. Other cosmological probes, such as strongly lensed fast radio bursts (FRBs), lensed gravitational-wave signals and so on would be beneficial complementary probes in the case $\Lambda \geq R_E$. Besides, the SIS model ($\gamma_0=2,\gamma_1=0$) is included within $1\sigma$ for all simulated samples selected with different criteria, which is consistent with the prior $\gamma=2.01 \pm 1.24$ used to model the total mass density of the forecasted LSST lenses.

In this work, we also considered four sub-samples derived from the well-defined sample of 5000 LSST lenses and applied the logarithmic polynomial cosmographic reconstruction of distances based on the SNe Ia pantheon sample. It should be noted that the slight deviation from $\gamma_{PN}=1$ ($\gamma_{PN}=0.981_{-0.020}^{+0.021}$ for  $\gamma=\delta$, and $\gamma_{PN}=0.935_{-0.131}^{+0.051}$ for  $\gamma \neq \delta$) is a bit more noticeable than in the case of the full simulated SGL sample ($\gamma_{PN}=0.998_{-0.007}^{+0.003}$ for $\gamma=\delta$, and $\gamma_{PN}=0.973_{-0.071}^{+0.027}$ for $\gamma \neq \delta$) in both lens models. This indicates the possibility that over the redshift range $0<z<2.5$, the SNe Ia pantheon sample serving as standard candles may provide a valuable supplement to the assumed fiducial cosmology in testing departure from GR. For the sub-samples defined by different redshift ranges, the constraints on $\gamma_{PN}$ became more diverse in the lens model where $\gamma \neq \delta$ is assumed. Significant departures from $\gamma_{PN}=1$ are present, which are $\gamma_{PN}=0.547_{-0.369}^{+0.346}$ and $\gamma_{PN}=0.546_{-0.375}^{+0.394}$ corresponding to the sub-samples $0.3 < z < 0.65$ and $z \geq 0.65$, respectively. However, in the case of $\gamma = \delta$, the same sub-samples gave $\gamma_{PN}=0.999_{-0.013}^{+0.007}$ and $\gamma_{PN}=0.999_{-0.097}^{+0.054}$ respectively. Therefore, the lens mass model seems to have a great influence on the limits on the PPN parameter with screening length $\Lambda=10 - 300 \; kpc$, which also can be concluded from the current intermediate mass SGL systems. Additionally, we have re-simulated the LSST samples with modified gravity effect present in the fiducial model and our results demonstrate the effectiveness of our methodology. In this paper, we just adopted a power-law profile to characterize the distribution of the luminous component. On the other hand, there are other more complicated but also more realistic descriptions of the luminosity density profiles in the literature \citep{Hernquist1990,Navarro1997}. It should be emphasized that more appropriate and accurate modeling of the structure of lensing galaxies will contribute to the precision and accuracy of testing the validity of GR using the SGL systems, and future systematic surveys such as LSST \citep{Abell2009}, DES \citep{Frieman2004}, and Euclid survey \citep{Pocino2021} will greatly conduce to such studies.

\vspace{0.5cm}

This work was supported by the National Natural Science Foundation of China under Grants Nos. 12021003, 11633001, and 11920101003; the Strategic Priority Research Program of the Chinese Academy of Sciences, Grant No. XDB23000000; the Interdiscipline Research Funds of Beijing Normal University; the China Manned Space Project (Nos. CMS-CSST-2021-B01 and CMS-CSST-2021-A01); and the CAS Project for Young Scientists in Basic Research under Grant No. YSBR-006. M.B. was supported by Foreign Talent Introducing Project and Special Fund Support of Foreign Knowledge Introducing Project in China (No. G2021111001L). He is also grateful for support from Polish Ministry of Science and Higher Education through the grant DIR/WK/2018/12.

\end{document}